\documentclass[journal,10pt]{IEEEtran}
\usepackage{amsmath, nccmath}
\usepackage{cases} 
\usepackage{amssymb}
\usepackage{cite}
\usepackage{url}
\usepackage{xcolor}
\usepackage{cite,graphicx}
\usepackage{subfigure}
\usepackage{citesort}
\usepackage{fancyhdr}
\usepackage{mdwmath}
\usepackage{mdwtab}
\usepackage{caption}
\usepackage{amsthm}
\usepackage{algorithmic}
\usepackage{algorithm}
\usepackage{threeparttable}
\usepackage{makecell}
\usepackage{booktabs}
\usepackage{pifont}
\newcommand{\cmark}{\ding{52}}

\newtheorem{remark}{Remark}
\newtheorem{theorem}{Theorem}

\newtheorem{lemma}{Lemma}

\newtheorem{corollary}{Corollary}

\makeatletter
\def\ScaleIfNeeded{%
\ifdim\Gin@nat@width>\linewidth \linewidth \else \Gin@nat@width
\fi } \makeatother

\begin{document}

\title{Pinching Antenna Systems for Integrated Sensing and Communications}

\author{
        {Haochen~Li},~\IEEEmembership{Member,~IEEE,} 
        {Ruikang~Zhong},~\IEEEmembership{ Member,~IEEE,}   
        {Zhiwen~Pan},~\IEEEmembership{Member,~IEEE,}
        {Chao~Dong},~\IEEEmembership{Senior Member,~IEEE,}
        {Jiayi~Lei},~\IEEEmembership{Member,~IEEE,}
        {Yuanwei~Liu},~\IEEEmembership{Fellow,~IEEE}   
\thanks{Part of this article has been submitted to the IEEE International Conference on Communications in 2026~\cite{li2025crb}.}
\thanks{Haochen~Li and Chao~Dong are with the College of Electronic and Information Engineering, Nanjing University of Aeronautics and Astronautics, Nanjing 211106, China (email: haochen.li@nuaa.edu.cn, dch@nuaa.edu.cn).}
\thanks{Ruikang~Zhong is with the School of Information and Communication Engineering, Xi'an Jiaotong University, Xi'an  710049, China (email: ruikang.zhong@xjtu.edu.cn).}
\thanks{Zhiwen~Pan is with National Mobile Communications Research Laboratory, Southeast University, Nanjing 210096, China, and also with Purple Mountain Laboratories, Nanjing 211100, China (email: pzw@seu.edu.cn).}
\thanks{Jiayi~Lei is with the School of Information and Communication Engineering, Beijing University of Posts and Telecommunications, Beijing 100876, China (e-mail: leijiayi@bupt.edu.cn).}
\thanks{Yuanwei Liu is with the Department of Electrical and Electronic Engineering, The University of Hong Kong, Hong Kong. (e-mail: yuanwei@hku.hk).}
}

\maketitle

\begin{abstract}
In this work, a multiple waveguide pinching antenna system (PASS) assisted integrated sensing and communication (ISAC) system is proposed, where the  base station (BS) is equipped with transmitting pinching antennas (PAs) and receiving uniform linear array (ULA) antennas. The PASS-transmitting-ULA-receiving (PTUR) BS transmits the communication and sensing signals through the  PAs on waveguides and collects the echo sensing signals with the mounted ULA. Based on this configuration, a target sensing Cramér–Rao Bound (CRB) minimization problem is formulated under communication quality-of-service (QoS) constraints, power budget constraint, and PA deployment constraints. To tackle the resulting non-convex problem, an alternating optimization (AO) framework is developed, which decomposes the problem into a digital beamforming sub-problem and a pinching beamforming sub-problem. The digital beamforming design is optimized via semidefinite relaxation (SDR), while the PA deployment is updated using penalty-based method. Simulation results demonstrate that: 1) the proposed PASS assisted ISAC framework achieves superior performance over benchmark schemes; and 2) the PASS assisted ISAC  is less affected by stringent communication constraints compared to conventional MIMO-ISAC, and benefits from increasing the number of waveguides and PAs per waveguide.



\end{abstract}


\section{Introduction}\label{I}
With the rapid evolution of wireless communication systems, advanced technologies such as multiple-input multiple-output (MIMO) and non-orthogonal multiple access (NOMA) have been developed to significantly enhance system capacity~\cite{9598915,6736761}. However, the continuous surge in capacity demand imposes increasingly stringent requirements on system design. To meet these challenges, several flexible antenna architectures—such as reconfigurable intelligent surfaces (RISs)\cite{9424177}, movable antennas (MAs)\cite{zhu2023movable}, and fluid antennas (FAs)\cite{wong2020fluid}—have recently attracted extensive attention from both academia and industry. These technologies reshape the long-standing paradigm of treating the wireless channels as uncontrollable, instead enabling active manipulation of the wireless channels. By introducing additional design degrees of freedom, flexible antenna systems offer new opportunities to improve overall system performance\cite{10945421}.

Along with growing capacity requirement, the evolution of communication systems also calls for the integration of multiple functionalities beyond data transmission. Among them, integrated sensing and communication (ISAC) has emerged as a promising paradigm, enabling the environmental awareness communications. Traditionally, communication and sensing functions were designed separately, leading to redundant hardware deployment and high operational costs~\cite{9737357}. ISAC, by contrast, leverages shared infrastructure and spectrum to jointly optimize both functionalities, thereby enhancing spectral efficiency while improving situational awareness~\cite{lu2024integrated}.

The integration of flexible antenna architectures into ISAC systems has been widely explored in recent studies~\cite{liu2023snr,li2025movable,10707252,10969546}. Despite their potential, several challenges persist, particularly line-of-sight (LoS) path blockage and severe propagation loss. For instance, RIS assisted ISAC can establish alternative LoS paths between the base station (BS) and sensing targets or communication users, but its passive nature inherently causes the well-known double-fading effect, resulting in significant propagation loss~\cite{basar2021reconfigurable}. By contrast, MA- and FA-assisted ISAC systems avoid double fading thanks to their active characteristics. Nevertheless, they remain highly susceptible to LoS blockages, as adjusting antennas within a limited wavelength-scale region cannot reliably generate unobstructed LoS links~\cite{wong2023fluid}.

To overcome these limitations, a novel flexible antenna architecture—namely the pinching antenna system (PASS)—has been recently proposed~\cite{yang2025pinching}. PASS deploys dielectric waveguides across the service area, on which dielectric particles, referred to as pinching antennas (PAs), can be flexibly applied~\cite{liu2025pinching}. On one hand, the wide coverage of waveguides and adjustable PA positions enable dynamic establishment of LoS paths to both users and sensing targets. On the other hand, since waveguides extend close to service locations, PAs can be positioned near users or targets, thereby effectively mitigating propagation loss, which typically scales with the square of transmission distance~\cite{suzuki2022pinching}.

Another distinctive advantage of PASS lies in its high reconfigurability. While the transceiver hardware architectures of MIMO, RIS, MA, and FA systems are typically fixed after deployment, PAs in PASS can be easily attached to or detached from the waveguides. This allows real-time reconfiguration in terms of both the number and placement of antennas. Consequently, PASS can adapt its antenna structure to dynamic communication and sensing environments in a scalable, efficient, and cost-effective manner~\cite{zeng2025resource}.

\subsection{Prior Works}
\subsubsection{PASS Assisted Communications} Although PASS systems provide unique advantages, their integration into existing communication frameworks and the design of efficient PA deployment strategies remain critical for fully unlocking their potential. A number of studies have investigated the design and application of PASS in communication systems. For instance, the authors of~\cite{10945421} highlighted the capacity enhancement and flexible deployment characteristics of PASS and proposed a PASS assisted NOMA system. Owing to the fact that multiple PAs mounted on the same waveguide share a common signal stream, PASS naturally facilitates the successive interference cancellation process inherent to NOMA. Analytical results confirmed that PASS can effectively mitigate large-scale path loss in wireless networks.

To further verify performance gains, the authors of~\cite{10896748} studied a simplified single-waveguide single-user scenario, optimizing PA placement to maximize communication rate. The results demonstrated the clear rate improvement enabled by PASS deployment. Extending this analysis, the authors of~\cite{hu2025sum} investigated a multiple-waveguide setup with one PA per waveguide under a PASS assisted NOMA framework. By jointly optimizing transmit beamforming and PA locations, the study revealed the critical role of in-waveguide power attenuation in determining system performance. More generally, the authors of~\cite{bereyhi2025mimo} considered multiple waveguides, multiple PAs, and multiple users, addressing both uplink and downlink design problems. Efficient algorithms were developed for joint transmit beamforming and PA deployment, demonstrating the scalability of PASS to complex network scenarios. Building upon these insights, researchers have further explored PASS assisted physical layer security\cite{wang2025pinching}, covert communications~\cite{jiang2025pinching}, multicast transmission~\cite{mu2025pinching}, and simultaneous wireless information and power transfer~\cite{li2025mimo}, confirming the versatility of PASS across diverse communication objectives.
\begin{table*}[t]

    \caption{{Comparisons with Previous PASS sensing Works}}\label{Comparisons4}
    \centering
    \begin{threeparttable}
    {{\begin{tabular}{*{11}{l}}
    \toprule
         & \makecell[c]{Our paper} & \cite{ding2025pinching} & \cite{wang2025wireless} & \cite{bozanis2025cram} & \cite{qin2025joint} & \cite{khalili2025pinching} & \cite{ouyang2025rate} & \cite{zhang2025integrated} & \cite{mao2025multi}   \\
    \midrule
    \midrule
        ISAC & \makecell[c]{\cmark} &   &  &    & \makecell[c]{\cmark} & \makecell[c]{\cmark} & \makecell[c]{\cmark} & \makecell[c]{\cmark} & \makecell[c]{\cmark}\\
    \midrule
        Sensing only & applicable & \makecell[c]{\cmark} &  \makecell[c]{\cmark}& \makecell[c]{\cmark} &   &  &   &  &   \\
    \midrule
        Single Waveguide & applicable & & \makecell[c]{\cmark} & \makecell[c]{\cmark} & \makecell[c]{\cmark} & \makecell[c]{\cmark}  & \makecell[c]{\cmark}  & \makecell[c]{\cmark} &   \\
    \midrule
        Multiple Waveguides & \makecell[c]{\cmark} & \makecell[c]{\cmark}  & &    &  &  &  &  & \makecell[c]{\cmark} \\
    \midrule
        Single Communication User & applicable &  &  &  & & \makecell[c]{\cmark} & \makecell[c]{\cmark} & \makecell[c]{\cmark} &   \\
    \midrule
        Multiple Communication Users & \makecell[c]{\cmark}  &   & & & \makecell[c]{\cmark} &  &  &  & \makecell[c]{\cmark} \\
    \midrule
        CRB & \makecell[c]{\cmark} & \makecell[c]{\cmark} &\makecell[c]{\cmark} & \makecell[c]{\cmark} & &  &  &  &   \\
    \midrule
        Sensing SINR/Power &  &  & &   & \makecell[c]{\cmark}&\makecell[c]{\cmark} &\makecell[c]{\cmark} & \makecell[c]{\cmark} & \makecell[c]{\cmark}  \\
    \bottomrule
    \end{tabular}}}
    \end{threeparttable}
    \end{table*}
\subsubsection{PASS Assisted ISAC} 
PASS has also been recognized as a promising architecture for enhancing wireless sensing and ISAC systems, owing to its reconfigurability, flexible deployment, and near-field channel characteristics. {\color{blue}The PASS assisted ISAC framework can be applied to practical scenarios such as smart manufacturing, intelligent transportation, and indoor sensing.} Several studies have investigated its role in different sensing configurations and objectives~\cite{ding2025pinching,wang2025wireless,bozanis2025cram,qin2025joint,khalili2025pinching,ouyang2025rate,zhang2025integrated,mao2025multi}.  For example, the authors of~\cite{ding2025pinching} analyzed the Cramér–Rao bound (CRB) in PASS assisted ISAC systems, showing that PAs can substantially improve positioning accuracy and achieve more uniform user performance. The study further highlighted the potential of PASS to support user-centric sensing due to its reconfigurability and low cost. Recognizing the complexity of using PAs for both transmission and echo collection, alternative hybrid designs have been proposed. The authors of~\cite{wang2025wireless} introduced a PASS and leaky coaxial (LCX) cable based sensing system, where PAs are used for transmission and LCX cables for reception. By jointly optimizing the transmit waveform and PA placement with a two-stage particle swarm optimization (PSO) algorithm, significant multi-target sensing accuracy gains were achieved. Similarly, the authors of~\cite{bozanis2025cram} examined a bistatic PASS assisted ISAC setup using a uniform linear array (ULA) for echo collection, deriving closed-form CRBs for joint range and angle estimation. Their results demonstrated that PASS can achieve high-resolution sensing with significantly fewer hardware resources compared to conventional arrays.

While existing studies~\cite{ding2025pinching,wang2025wireless,bozanis2025cram} validate the potential of PASS in sensing, they are  limited to \textit{sensing-only} scenarios. The adoption of PASS in ISAC requires carefully balancing communication and sensing objectives under complex design constraints. To this end, the authors of~\cite{qin2025joint} formulated a joint PA placement and power allocation problem, aiming to maximize communication rate while satisfying sensing and energy constraints, and solved it with a maximum entropy reinforcement learning algorithm. Superior performance was reported in terms of data rate, sensing SNR, and robustness. In study~\cite{khalili2025pinching}, a novel PASS assisted ISAC architecture was proposed, where dynamically activated PAs achieve target diversity. By modeling radar cross section as a random variable and adopting outage probability as a reliability metric, the PA activation problem was solved using a successive convex approximation (SCA) method. The authors of~\cite{ouyang2025rate} analyzed the fundamental limits of PASS assisted ISAC systems, deriving closed form expressions for achievable communication and sensing rates under various beamforming objectives, including sensing centric, communication centric, and Pareto optimal designs. The authors of~\cite{zhang2025integrated} optimized PA mobility for illumination power maximization subject to user QoS constraints, developing a penalty based alternating optimization (AO) algorithm. Most recently, the authors of~\cite{mao2025multi} studied a multi-waveguide PASS assisted ISAC framework, where transmit and receive PAs were jointly optimized under rate, SNR, power, and placement constraints. A fine tuning approximation and SCA based method were proposed, demonstrating the feasibility of large scale PASS deployment in practical ISAC systems.

\subsection{Motivations and Contributions}


The aforementioned PASS assisted ISAC studies~\cite{qin2025joint,khalili2025pinching,ouyang2025rate,zhang2025integrated,mao2025multi} have employed \textit{sensing SNR} or \textit{sensing power} as performance metrics. These metrics are attractive owing to their computational simplicity and their direct dependence on hardware and system parameters such as transmit power, beamforming gain, and antenna configuration. Nevertheless, sensing SNR does not explicitly capture the accuracy of parameter estimation~\cite{wei2023integrated}.

In contrast, the CRB provides a theoretical lower bound on the variance of unbiased estimators, thereby serving as a fundamental benchmark for estimation accuracy. CRB is independent of specific signal processing algorithms and can accommodate multi-parameter estimation scenarios~\cite{9540344}. To the best of our knowledge, no prior work has investigated the CRB for target sensing in PASS assisted ISAC systems. To fill this gap, this paper adopts CRB as the sensing performance metric in a PASS assisted ISAC system that simultaneously serves multiple communication users and senses a target, thereby offering a more fundamental and algorithm-independent evaluation of sensing accuracy. A comparison with previous PASS sensing works is provided in Table~\ref{Comparisons4}. 
The main contributions of this work can be summarized as follows:
\begin{itemize}
        \item A general multiple waveguide PASS assisted ISAC system with the PASS-transmitting-ULA-receiving (PTUR) BS is proposed. A CRB minimization problem is formulated and then decomposed into a digital beamforming sub-problem and a pinching beamforming sub-problem with the AO method. By solving the formulated optimization problem, the minimum CRB for PASS assisted ISAC under given power budget and communication QoS constraints is obtained for the first time, providing a fundamental benchmark for sensing accuracy and offering valuable insights into the trade-offs between sensing and communication performance.
        \item For the digital beamforming sub-problem with fixed PA deployment, the semidefinite relaxation (SDR) based optimization framework is developed to solve the non-convex sensing CRB minimization problem with rank-one constraint.  Optimal rank-one solutions can be constructed from the relaxed solution without performance loss, guaranteeing globally optimal beamforming and sensing covariance matrix design.
        \item For the pinching beamforming sub-problem with fixed digital beamforming and sensing covariance matrices, an AO framework is developed, combining penalty, SCA, and element-wise optimization techniques to address the non-convex and tightly coupled PA deployment constraint. By introducing auxiliary variables, the interactions between the communication channels and the sensing CRB are decoupled. The penalty method is employed to handle the constraints associated with the auxiliary variables, SCA is used to tackle the non-convex signal-to-interference-plus-noise ratio (SINR) constraints, and a one-dimensional search is applied for element-wise PA position updates, ensuring efficient and convergent optimization.

        \item The simulation results demonstrate that: 1) the proposed PASS assisted ISAC framework consistently outperforms other benchmark schemes, which verifies the effectiveness of proposed algorithms; 2) PASS assisted ISAC schemes exhibit stronger tolerance against performance degradation under stringent communication constraints compared to conventional MIMO-ISAC schemes, highlighting the benefits introduced by pinching beamforming; and 3) increasing the number of waveguides as well as the number of PAs per waveguide further enhances the efficiency of pinching beamforming, thereby improving overall system performance.
\end{itemize}

\subsection{Organization and Notations}
The remainder of this paper is organized as follows. In Section~\ref{II}, we present the system model, including the proposed PASS architecture and associated channel and signal models. Section~\ref{III} formulates the joint optimization problem and introduces the AO algorithm. Section~\ref{sec:results} provides simulation results to evaluate the performance of the proposed system. Finally, Section~\ref{sec:conclusions} concludes the paper.

Notations: Boldface lowercase letters  denote vectors, boldface uppercase letters denote matrices, and normal letters denote scalars. The superscripts $\left(\cdot\right)^\mathrm{T}$ and $\left(\cdot\right)^\mathrm{H}$ represent transpose and Hermitian transpose, respectively. The notation $\|\cdot\|_2$ represents the Euclidean norm, $\|\cdot\|_F$ represents the Frobenius norm, $|\cdot|$ represents the modulus of a complex number, and $\text{tr}(\cdot)$ denotes the trace of a matrix. $\mathcal{CN}(\mathbf{a},\mathbf{A})$ denotes a complex Gaussian distribution with mean vector $\mathbf{a}$ and covariance matrix $\mathbf{A}$. The symbol $\mathbf{A}\succeq0$  indicates that the matrix $\mathbf{A}$ is positive semidefinite.  $\left[\mathbf{A}\right]_{mn}$ denotes the  element on the $m$-th row and $n$-th column of matrix $\mathbf{A}$. 
$\mathcal{O}\left(\cdot\right)$ represents the big-O notation, used to characterize the computational complexity of an algorithm. 
$\Re\left(\cdot\right)$ and $\Im\left(\cdot\right)$ denote the real and imaginary parts of a complex quantity, respectively. 
$ \mathbb{E}\left\{\cdot\right\}$ denotes the expectation operator. $\text{Bdiag}\left(\cdot\right) $denotes a block-diagonal matrix constructed from its arguments.

\section{System Model}\label{II}
\begin{figure} [htbp]
\centering
\includegraphics[width=0.5\textwidth]{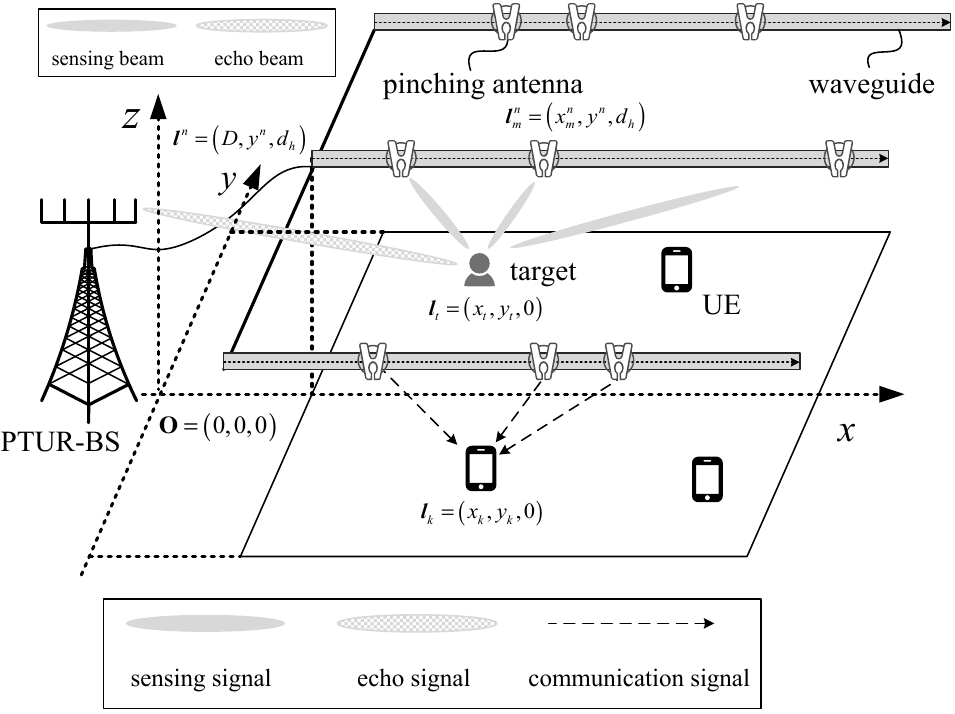}
 \caption{\color{blue}{The proposed PASS assisted ISAC system.}
  }
 \label{system_model}
\end{figure}
As shown in Fig.~\ref{system_model}, we consider a  PASS assisted  ISAC system, where the BS adopts the PTUR architecture, i.e., the signals are transmitted with a pinching antenna and the sensing echo is received with a ULA\footnote{{\color{blue}Employing a conventional ULA at the receiver, instead of a PASS  simplifies the overall system design and implementation}}. The dual function BS simultaneously communicates with $K$ user  and probes a target in the service area denoted by $\mathcal{A}$ with a $D_x \times D_y$ rectangle shape. The PASS is equipped with $N$ waveguides and $M$ pinching antennas are deployed on each waveguide, while the ULA is equipped with $M_r$ antennas. {\color{blue}To fully explore the potential of the proposed system, the waveguide-multiplexing architecture is adopted in this work instead of the waveguide-division counterpart~\cite{liu2025pinching}.} Let~$\mathcal{M}$,~$\mathcal{N}$, $\mathcal{M}_r$, and $\mathcal{K}$ denote the sets of all PAs on one waveguide, all waveguides, all ULA antenna elements, and all communication users, respectively.

All UEs and the target are uniformly distributed in $\mathcal{A}$, and  the location of the $k$-th UE and the target are $\boldsymbol{l}_k=\left[x_k,y_k,0\right]^\mathrm{T}, \forall k\in \mathcal{K},$ and $\boldsymbol{l}_t=\left[x_t,y_t,0\right]^\mathrm{T}$,respectively.   All waveguides are elevated at height $d_h$ while extending along the $x$-aixs, and the fed point of the $n$-th waveguide is $\boldsymbol{l}^n=\left[D,y^n,d_h\right]^\mathrm{T}, \forall n\in \mathcal{N}$. The location of the $m$-th pinching antenna on the $n$-th waveguide is $\boldsymbol{l}_m^n=\left[x_m^n,y^n,d_h\right]^\mathrm{T}, \forall n\in \mathcal{N}, \forall m\in \mathcal{M}$. In PASS, the overall antenna architecture can be reconfigured by relocating any PA to a desired position, which accordingly updates the parameter $x_m^n$ corresponding to the $m$-th PA on the $n$-th waveguide. Generally, the position of each PA is subject to the following two constraints. First, the PAs have to remain on the waveguide, leading to the condition $0 \le x_m^n \le L$, $\forall n\in \mathcal{N}, \forall m\in \mathcal{M}$, where $L$ denotes the length of the waveguide. Second, to avoid mutual coupling effects between PAs, the distance between any two adjacent PAs on the same waveguide should exceed a minimum separation $\delta$, resulting in the constraint $x_m^n - x_{m-1}^n \ge \delta$, $\forall n\in \mathcal{N}, \forall m\in \mathcal{M}$. Let $\mathcal{X}$ denotes the $x$-aixs parameters for all PAs, it can be expressed as
\begin{equation}\label{X}
\begin{aligned}
\mathcal{X}=\left\{ {\begin{array}{*{20}{c}}
{x_m^n,}&\vline& {D \le x_m^n \le D+L}\\
{ \forall n, m}&\vline& {x_m^n - x_{m-1}^n \ge \delta}
\end{array}} \right\}.
\end{aligned}
\end{equation}
{\color{blue}The PAs can be mechanically attached to or detached from different points along the waveguide, which allows flexible spatial reconfiguration~\cite{liu2025pinching1}. Considering the fabrication complexity of fully continuous movement, a discrete PASS implementation with multiple predefined attachment slots can also be employed in practice. The continuously adjustable PA positions considered in this work thus serve as an idealized model to facilitate analytical tractability and reveal the upper-bound performance of the proposed system.}


\subsection{PASS Channel Model} 
The channels of PASS systems contain both the  wireless PA-UE channels and the in-waveguide channels. In the following, we introduce the  wireless channels in the PASS systems first.  Thanks to the flexible deployment option of PAs, the distance between PAs is significantly large than that of conventional antenna arrays (typically sub wavelength). Thus, all UEs served by the PAs are located in their near-field region, where the propagation of electromagnetic waves should be accurately described with  spherical waves\footnote{{\color{blue}For example, consider a PASS with an array diameter of $D=5$ m operating at a frequency of $f=28$ GHz. In this case, the corresponding Rayleigh distance can be calculated as $d_{\text{Rayleigh}} = {2D^2f}/{c} \approx 4.67$~km.}}. In this case, the wireless channel between the PAs on the $n$-th waveguide and user $k$ can be expressed as
\begin{equation}
\mathbf{h}^n_k=\left(2\kappa_c\right)^{-1}\left[\frac{e^{-j \kappa_c r_{k,1}^n}} {r_{k,1}^n},\frac{e^{-j \kappa_c r_{k,2}^n}} {r_{k,2}^n},\cdots,\frac{e^{-j \kappa_c r_{k,M}^n}} {r_{k,M}^n}\right]^\mathrm{T},
\end{equation}
where $\kappa_c={2\pi f_c}/{c}={2\pi}/{\lambda_c}$ is the wavenumber. $r_{k,m}^n=\left\|\boldsymbol{l}_m^n-\boldsymbol{l}_k\right\|_2$ stands for the distance between the $m$-th PA on the $n$-th waveguide and user $k$. Then, the overall wireless channel between the PAs and user $k$ can be expressed as
\begin{equation}
\hat{\mathbf{h}}_k=\left[\left(\mathbf{h}^1_k\right)^\mathrm{T},\left(\mathbf{h}^2_k\right)^\mathrm{T},\cdots,\left(\mathbf{h}^N_k\right)^\mathrm{T}\right]^\mathrm{T}\in\mathbb{C}^{MN\times 1}.
\end{equation}
Similarly, the wireless channel between the PAs on the $n$-th waveguide and target can be given as
\begin{equation}
\mathbf{h}^n_t=\left(2\kappa_c\right)^{-1}\left[\frac{e^{-j \kappa_c r_{t,1}^n}} {r_{t,1}^n},\frac{e^{-j \kappa_c r_{t,2}^n}} {r_{t,2}^n},\cdots,\frac{e^{-j \kappa_c r_{t,M}^n}} {r_{t,M}^n}\right]^\mathrm{T},
\end{equation}
where $r_{t,m}^n=\left\|\boldsymbol{l}_m^n-\boldsymbol{l}_t\right\|_2$ stands for the distance between the $m$-th PA on the $n$-th waveguide and target. Then, the wireless channel between the PAs and the target can be expressed as
\begin{equation}
\hat{\mathbf{h}}_t=\left[\left(\mathbf{h}^1_t\right)^\mathrm{T},\left(\mathbf{h}^2_t\right)^\mathrm{T},\cdots,\left(\mathbf{h}^N_t\right)^\mathrm{T}\right]^\mathrm{T}\in\mathbb{C}^{MN\times 1}.
\end{equation}
{\color{blue}The wireless channels in PASS systems are modeled as  LoS  channels. This assumption is justified for two main reasons. First, the positions of the  PAs  can be flexibly adjusted along the waveguides, and with careful PASS deployment design, LoS links between the PAs and users can generally be maintained. Second, PASS assisted systems typically operate in high-frequency bands (28 GHz in this work), where LoS propagation usually dominates the channel characteristics.}

Then, the in-waveguide channel of the PASS system is introduced. On one hand, the in-waveguide channel describes the electromagnetic wave propagation delay and attenuation inside the waveguide. Since the propagation attenuation inside the waveguide is much smaller than that in the air, the propagation attenuation is neglected in this work and we focus on the influence of the propagation delay introduced by the in-waveguide transmission. On the other hand, the structure of the in-waveguide channel reflects the structure of the PAs, as each PA can only be attached to one specific waveguide. Thus, the in-waveguide channel is in the form of a block-diagonal matrix, which can be expressed as 
\begin{equation}
\begin{aligned}
\mathbf{F}&=\text{Bdiag}\left(\mathbf{f}_1,\mathbf{f}_2,\cdots,\mathbf{f}_N\right)^\mathrm{T}\in\mathbb{C}^{N\times MN}
\end{aligned}
\end{equation}
where $\mathbf{f}_n=\left[f_1^n,f_2^n,\cdots,f_M^n\right]^\mathrm{T}\in\mathbb{C}^{M\times 1}$ with $f_m^n$ stands for the phase difference between the signal transmitted from the $m$-th PA of the $n$-th waveguide and the in-put signal for the $n$-th waveguide. Specifically, it can be expressed as 
\begin{equation}
f_m^n = \rho_m^ne^{-j\kappa_gd_m^n},
\end{equation}
where $\kappa_g={2\pi}/{\lambda_g}$ is the guided wavenumber, and $\lambda_g=\lambda_c/n_{e}$ is the guided wavelength of the waveguide with effective refractive index $n_{e}$. $d_{m}^n=\left\|\boldsymbol{l}^n-\boldsymbol{l}_m^n\right\|_2$ stands for the distance between the $m$-th PA on the $n$-th waveguide and  the fed point of the $n$-th waveguide. $\rho_m^n$ denotes the power allocation coefficient of the $m$-th PA on the $n$-th waveguide. {\color{blue}For simplicity, the equally power allocation strategy is adopted within each waveguide, i,e., $\rho_m^n=\rho=1/\sqrt{M}, \forall m,n$. The in-waveguide propagation loss is negligible compared with free-space path loss and is thus omitted for simplicity, while incorporating it in future designs would lead to a more accurate and practical channel model.}  Then, the overall channel between the PASS and user $k$ can be given as 
\begin{equation}\label{h_k}
\mathbf{h}_k=\mathbf{F}\hat{\mathbf{h}}_k\in\mathbb{C}^{N\times 1}.
\end{equation}
The overall channel between the PASS and the target can be given as
\begin{equation}
\mathbf{h}_t=\mathbf{F}\hat{\mathbf{h}}_t\in\mathbb{C}^{N\times 1}.
\end{equation}
{\color{blue}The channel state information of the PASS assisted systems can be obtained using channel estimation method in~\cite{zhou2025channel}.}
\subsection{Signal Model} 
During downlink ISAC, the PAs send communication and sensing signals to simultaneously serve all communication UEs and probe the target. The echo signals reflected by the target is collect by the ULA of the BS. At time slot $t$, the PA transmission signal can be expressed as 
\begin{equation}
\mathbf{s}\left(t\right)=\mathbf{W}\mathbf{c}\left(t\right)+\mathbf{c}_s\left(t\right),
\end{equation}
where $\mathbf{W}=\left[\mathbf{w}_1,\mathbf{w}_2,\cdots,\mathbf{w}_K\right]\in\mathbb{C}^{N\times K}$ is the beamforming matrix for communication users, with $\mathbf{w}_k\in\mathbb{C}^{N\times 1}$ denoting the beamformer assigned to user $k$, $\forall k \in \mathcal{K}$. $\mathbf{c}\left(t\right)=\left[{c}_1\left(t\right),{c}_2\left(t\right),\cdots,{c}_K\left(t\right)\right]^\mathrm{T}\in\mathbb{C}^{K\times 1}$ is the communication signal transmitted at time slot $t$, with $c_k\left(t\right)\in\mathbb{C}$ denoting the signal for user $k$, $\forall k \in \mathcal{K}$. The communication signals  for different users are independent, thus the normalized communication signal satisfies $\mathbb{E}\left\{\mathbf{c}\left(t\right)\mathbf{c}\left(t\right)^\mathrm{H}\right\}=\mathbf{I}_K$. $\mathbf{c}_s$ is the dedicate sensing signal with covariance matrix $\mathbf{R}_s$. The covariance matrix of the PA transmission signal is 
\begin{equation}
        \mathbf{R}=\mathbb{E}\left\{\mathbf{s}\left(t\right)\mathbf{s}\left(t\right)^\mathrm{H}\right\}=\mathbf{W}\mathbf{W}^\mathrm{H}+\mathbf{R}_s.
\end{equation}

\subsection{Communication Model and Sensing Model}
At time slot $t$, the received signal of user $k$ can be expressed as 
\begin{equation}
\begin{aligned}
        {y}_k\left(t\right)=\mathbf{h}_k^\mathrm{H}\mathbf{w}_k{c}_k\left(t\right)+\sum\nolimits_{i\ne k}\mathbf{h}_k^\mathrm{H}\mathbf{w}_i{c}_i\left(t\right)+\mathbf{h}_k^\mathrm{H}\mathbf{c}_s\left(t\right)+n_k\left(t\right),
\end{aligned}
\end{equation}
where ${n}_k\left(t\right) \sim \mathcal{CN}({0}, \sigma_0^2)$ denotes the additive white Gaussian noise at the receiver of user $k$. Treating the received sensing signal as interference, the communication SINR can be given as
\begin{equation}\label{CommunicationSINR}
        \text{SINR}_{k} = \frac{|\mathbf{h}_{k}^\mathrm{H}\mathbf{w}_{k}|^2}{\sum\nolimits_{i\neq k}|\mathbf{h}_{k}^\mathrm{H}\mathbf{w}_{i}|^2+\mathbf{h}_{k}^\mathrm{H}\mathbf{R}_{s}\mathbf{h}_{k}+\sigma_0^2}.
\end{equation}

To carry out the sensing function, the BS sends probing signals with PAs and collects the reflected signals with the ULA. The echo signal received by the BS ULA at time slot $t$ is 
\begin{equation}      
        \mathbf{y}_s\left(t\right)=\mathbf{G}\mathbf{s}\left(t\right)+\mathbf{H}_\mathrm{I}\mathbf{s}\left(t\right)+{\mathbf{n}}_s\left(t\right),
\end{equation}
where ${\mathbf{n}}_s\left(t\right)\sim \mathcal{CN}(\mathbf{0}, {\sigma}_s^2\mathbf{I}_{M_r})$ denotes the additive white Gaussian noise vector at the receiving ULA. $\mathbf{H}_\mathrm{I}\in\mathbb{C}^{M_r\times MN}$ stands for the self-interference channel between the PAs and the receiving ULA. $\mathbf{G}$ is the PASS-target-ULA two-hop sensing channel. It can be  expressed as 
\begin{equation}
        \mathbf{G}=\beta\mathbf{a}\left(\theta\right)\mathbf{h}_t^\mathrm{H},
\end{equation}
where $\beta$ incorporates target reflection factor and the path loss between the BS ULA and target. $\mathbf{a}\left(\theta\right)$ is the steering vector of the $M_r$-element ULA with half wavelength element distance, i.e.,
\begin{equation}
    \mathbf{a}\left(\theta\right)=[1,e^{-jk_c \sin\theta},\cdots,e^{-jk_c\left(M_r-1\right)\sin\theta}]^\mathrm{T},    
\end{equation}
where $\theta$ represents the angle of the target with respect to the receiving ULA. 
\begin{remark}
{\color{blue}\emph{The proposed PASS assisted ISAC system with the PTUR architecture can effectively mitigate self-interference. The main reasons are as follows. {\textit{Firstly}}, compared with the monostatic MIMO ISAC system, the PAs in PASS are placed farther away from the BS receiving antennas, which introduces additional attenuation to the leakage signals. Moreover, the increased physical separation between the transmitting and receiving antennas creates favorable conditions for adopting more effective isolation methods. {\textit{Secondly}}, since the waveguides in PASS are deployed at a certain height while the sensing targets are located in the lower service area, the sensing echoes and the self-interference signals of PASS arrive at the receiver from different incident angles. This angular diversity enables the BS to apply angle domain filtering to preserve the desired sensing echoes while suppressing the residual self-interference signals. }}
\end{remark}

Owing to the above mitigation mechanisms, the impact of self-interference is significantly reduced in the proposed PASS assisted ISAC system. Therefore, self-interference is neglected in this work. The echo signal received by the BS ULA at time slot $t$ is 
\begin{equation}      
        \mathbf{y}_s\left(t\right)=\mathbf{G}\mathbf{s}\left(t\right)+\mathbf{n}_s\left(t\right).
\end{equation}

\subsection{Sensing CRB Derivation}
The BS collects the echo signals over $T$ time slot, which can be expressed as 
\begin{equation}
        \mathbf{Y}_s=\mathbf{G}\mathbf{S}+\mathbf{N}_s,
\end{equation}
where $\mathbf{Y}_s=[ \mathbf{y}_s\left(1\right),\mathbf{y}_s\left(2\right), \cdots , \mathbf{y}_s\left(T\right)]$, $\mathbf{S}=[ \mathbf{s}\left(1\right),\mathbf{s}\left(2\right), \cdots , \mathbf{s}\left(T\right)]$, and $\mathbf{N}_s=[ \mathbf{n}_s\left(1\right),\mathbf{n}_s\left(2\right), \cdots , \mathbf{n}_s\left(T\right)]$.

Define the parameter vector as 
\begin{equation}
        \boldsymbol{\eta} = \left[x_t,y_t,\Re\left(\beta\right),\Im\left(\beta\right)\right]^\mathrm{T}\in\mathbb{R}^{4\times 1}.
\end{equation} 
The goal of wireless sensing is to estimate parameters in $\eta$ with received signal $\mathbf{Y}_s$. Mean square error (MSE) is generally adopted as the metric of estimation and the Cram\'er-Rao bound can give the MES lower bound of any unbiased estimator. The FIM of $\boldsymbol{\eta}$ is 
\begin{equation}\label{eqn:FIM}
        \mathbf{J}=\begin{bmatrix}
         \mathbf{J}_{11} & \mathbf{J}_{12} \\
         \mathbf{J}_{12}^\mathrm{T} & \mathbf{J}_{22} 
    \end{bmatrix}\in\mathbb{R}^{4\times 4},
\end{equation}
where $\mathbf{J}_{11}$ is the sub FIM with respect to target location parameters $x_t$ and  $y_t$. $\mathbf{J}_{22}$ is the sub FIM with respect to $\beta$. $\mathbf{J}_{12}$ is the sub FIM with respect to the coupling between target location parameters and $\beta$. The detailed expressions of ${\bf{J}}_{11}$, ${\bf{J}}_{12}$, and ${\bf{J}}_{22}$ are derived in Appendix~A. The CRB of estimating target location $x_t$ and  $y_t$ can be given as 
\begin{equation}\label{eqn:CRB}
    \text{CRB}\left(x_t, y_t\right) = \left({\bf{J}}_{11}-{\bf{J}}_{12}{\bf{J}}_{22}^{-1}{\bf{J}}_{12}^T\right)^{-1}\in \mathbb{R}^{2 \times 2}.
\end{equation}
\subsection{Problem Formulation}
The objective is to minimize the trace of the CRB matrix, which characterizes the lower bound on the localization MSE of the target position $(x_t, y_t)$. The proposed PASS assisted ISAC system is designed to achieve high-accuracy target sensing performance while simultaneously meeting the QoS requirements of all communication users under a limited power budget. The corresponding optimization problem is formulated as follows:
\begin{subequations}\label{problem:CRB}
    \begin{align}        
        \min_{\mathbf{X},\mathbf{W}, \mathbf{R}_s} \quad &  \text{tr}\left(\text{CRB}\left(x_t, y_t\right)\right) \\
        \label{constraint:PASS}
        \mathrm{s.t.} \quad & \left[\mathbf{X}\right]_{mn} \in \mathcal{X}, \forall m \in \mathcal{M},n\in \mathcal{N},  \\ 
        \label{constraint:SINR}
        & \text{SINR}_{k} \ge \gamma_{k}, \forall k \in \mathcal{K},\\
        \label{constraint:power}
        & \text{tr}\left(\mathbf{W}\mathbf{W}^H+\mathbf{R}_{s}\right) \le P,\\ 
        \label{constraint:semidefinite}
        & \mathbf{R}_{s}  \succeq  0,
    \end{align}
\end{subequations}
where $\mathbf{X}$ is the PA deployment variable matrix, whose entry on the $m$-th row and $n$-th column is $x_m^n$. Constraint~\eqref{constraint:PASS} ensures that the placement of PAs complies with the hardware and electromagnetic limits defined in~\eqref{X}.
Constraint~\eqref{constraint:SINR} guarantees the communication  QoS  requirements for all users in the set $\mathcal{K}$.
Constraint~\eqref{constraint:power} imposes a total transmit power budget $P$ for the PASS system.
Finally, constraint~\eqref{constraint:semidefinite} ensures the positive semi-definiteness of the sensing covariance matrix $\mathbf{R}_d$.

\section{Proposed Alternating Optimization Algorithm}\label{III}

To efficiently solve the non-convex problem in \eqref{problem:CRB}, we propose an AO algorithm. Specifically, the original problem is decomposed into two sub-problems: the  digital beamforming problem jointly optimizes the communication beamforming matrix $\mathbf{W}$ and the sensing covariance matrix $\mathbf{R}_s$, while the pinching beamforming problem focuses on optimizing the PA deployment $\mathbf{X}$. At each iteration, one set of variables is updated while keeping the others fixed, and this process continues until convergence.

\subsection{Optimize $\mathbf{W}$  and $\mathbf{R}_s$ with Fixed $\mathbf{X}$}\label{subsection_III_A}

When the PA positions $\mathbf{X}$ are fixed, the sensing and communication channels are known. The optimization problem becomes
\begin{subequations}\label{problem:CRB_WR}
    \begin{align}        
        \min_{ \mathbf{W}, \mathbf{R}_s} \quad &  \text{tr}\left(\text{CRB}\left(x_t, y_t\right)\right) \\
        \mathrm{s.t.} \quad & \eqref{constraint:SINR}\sim\eqref{constraint:semidefinite}, 
    \end{align}
\end{subequations}
This digital beamforming sub-problem is  non-convex due to the highly non-convex CRB expression in the objective function and non-convex constraint~\eqref{constraint:SINR}. To further simplify the  sub-problem, introduce an auxiliary semi-definite matrix $\mathbf{U}$ which follows
\begin{equation}
        \text{tr}(\mathbf{U}^{-1})\ge\text{tr}\left(\text{CRB}\left(x_t, y_t\right)\right)=\text{tr}(({\bf{J}}_{11}-{\bf{J}}_{12}{\bf{J}}_{22}^{-1}{\bf{J}}_{12}^T)^{-1}).
\end{equation}
Note that the function $\text{tr}(\mathbf{M}^{-1})$ is monotonic over any semi-definite matrix $\mathbf{M}$, thus we have $\left({\bf{J}}_{11}-\mathbf{U}\right)-{\bf{J}}_{12}{\bf{J}}_{22}^{-1}{\bf{J}}_{12}^T\succeq 0$, which can be expressed as following linear matrix inequality
\begin{equation}\label{eqn:LIM}
        \begin{bmatrix}
         \mathbf{J}_{11}-\mathbf{U} & \mathbf{J}_{12} \\
         \mathbf{J}_{12}^\mathrm{T} & \mathbf{J}_{22} 
    \end{bmatrix}\succeq 0.
\end{equation}
The problem~\eqref{problem:CRB_WR} can be reformulated as
\begin{subequations}\label{problem:CRB_WR_U}
    \begin{align}        
        \min_{\mathbf{U}, \mathbf{W}, \mathbf{R}_s} \quad &  \text{tr}\left(\mathbf{U}^{-1}\right) \\
        \mathrm{s.t.} \quad &~\eqref{constraint:SINR},~\eqref{constraint:power},~\eqref{eqn:LIM},\\ 
        & \mathbf{R}_{s}, \mathbf{U}  \succeq  0.
    \end{align}
\end{subequations}
The reformulated problem is still  non-convex due to the  non-convex constraint~\eqref{constraint:SINR}. This constraint can be rewritten as 
\begin{equation}\label{RW_SINR}
        \frac{1}{\gamma_{k}}|{\mathbf{h}}_k^\mathrm{H}{\mathbf{w}}_{k}|^2 \ge \sum\nolimits_{i\ne k}|{\mathbf{h}}_k^\mathrm{H}{\mathbf{w}}_{i}|^2+{\mathbf{h}}_k^\mathrm{H}{\mathbf{R}}_{s}{\mathbf{h}}_k+\sigma_0^2. 
\end{equation}
Define ${\mathbf{W}}_k={\mathbf{w}}_k{\mathbf{w}}_k^\mathrm{H}$, $\forall k \in \mathcal{K}$. Constraint~\eqref{constraint:SINR} can be further rewritten as 
\begin{equation}\label{RW_SINR_matrix}
\begin{aligned}
     \frac{1+\gamma_{k}}{\gamma_{k}}&\text{tr}\left({\mathbf{h}}_k{\mathbf{h}}_k^\mathrm{H}{\mathbf{W}}_{k}\right)\ge\sum\nolimits_{i}\text{tr}\left({\mathbf{h}}_k{\mathbf{h}}_k^\mathrm{H}{\mathbf{W}}_{i}\right) \\ &+\text{tr}\left({\mathbf{h}}_k{\mathbf{h}}_k^\mathrm{H}{\mathbf{R}}_{s}\right)+\sigma_0^2=\text{tr}\left({\mathbf{h}}_k{\mathbf{h}}_k^\mathrm{H}{\mathbf{R}}\right)+\sigma_0^2.   
\end{aligned}
\end{equation}
The constraint~\eqref{constraint:power} can be reformulated as
\begin{equation}\label{constraint:power_matrix} 
        \text{tr}\left(\sum\nolimits_{k=1}^K\mathbf{W}_k+\mathbf{R}_{s}\right) \le P
\end{equation}
Substitute~\eqref{RW_SINR_matrix},~\eqref{constraint:power_matrix}, and auxiliary matrices ${\mathbf{W}}_k, \forall k $ into problem~\eqref{problem:CRB_WR_U}, the  reformulated problem can be given as
\begin{subequations}\label{problem:CRB_WR_U_matrix}
    \begin{align}        
        \min_{\mathbf{U}, \left\{\mathbf{W}_k\right\}_{k=1}^K, \mathbf{R}_s} \quad &  \text{tr}\left(\mathbf{U}^{-1}\right) \\
        \mathrm{s.t.} \quad &~\eqref{eqn:LIM},~\eqref{RW_SINR_matrix},~\eqref{constraint:power_matrix}, \\\label{constraint:semidefinite_all}
        & \mathbf{R}_{s}, \mathbf{U}, \mathbf{W}_k, \forall k,  \succeq  0,\\
        \label{constraint:rank}
        & \text{rank}\left(\mathbf{W}_k\right)=1, \forall k.
    \end{align}
\end{subequations}
This optimization problem is non-convex due to the non-convex rank-one constraint~\eqref{constraint:rank}. By relaxing this constraint, the SDR of problem~\eqref{problem:CRB_WR_U_matrix} is
\begin{subequations}\label{problem:CRB_WR_U_matrix_SDR}
    \begin{align}        
        \min_{\mathbf{U}, \left\{\mathbf{W}_k\right\}_{k=1}^K, \mathbf{R}_s} \quad &  \text{tr}\left(\mathbf{U}^{-1}\right) \\
        \mathrm{s.t.} \quad &~\eqref{eqn:LIM},~\eqref{RW_SINR_matrix},~\eqref{constraint:power_matrix},~\eqref{constraint:semidefinite_all}.
    \end{align}
\end{subequations}
This optimization problem is convex and can be efficiently solved using CVX~\cite{grant2014cvx}. The computational complexity of solving problem~\eqref{problem:CRB_WR_U_matrix_SDR}  can be expressed as $\mathcal{O}\left(\left(K+1\right)^{4.5}N^{4.5}\right)$~\cite{5447068}.  Suppose $\{\{\hat{\mathbf{W}}_k\}_{k=1}^K, \hat{\mathbf{R}}_s\}$ is the solution to problem~\eqref{problem:CRB_WR_U_matrix_SDR}. $\{\hat{\mathbf{W}}_k\}_{k=1}^K$ obtained from the SDR generally has a  rank that is higher than one.

It has been proved that a set of solution $\{\{\bar{\mathbf{W}}_k\}_{k=1}^K, \bar{\mathbf{R}}_s\}$ that satisfies constraint~\eqref{constraint:rank} always can be constructed from the optimal solution of problem~\eqref{problem:CRB_WR_U_matrix_SDR}, i.e., $\{\{\hat{\mathbf{W}}_k\}_{k=1}^K, \hat{\mathbf{R}}_s\}$~\cite{9124713}. Specifically, for each user $k$, let $\hat{\mathbf{w}}_k$
\begin{equation}
        \bar{\mathbf{w}}_k=\left({\mathbf{h}}_k^\mathrm{H}\hat{\mathbf{W}}_k{\mathbf{h}}_k\right)^{-1}\hat{\mathbf{W}}_k{\mathbf{h}}_k.
\end{equation}
The constructed rank-one solution is $\bar{\mathbf{W}}_k=\bar{\mathbf{w}}_k\bar{\mathbf{w}}_k^\mathrm{H}$. Then the sensing covariance matrix can be calculated as 
\begin{equation}
     \bar{\mathbf{R}}_s= \hat{\mathbf{R}}_s- \sum\nolimits_{i}\bar{\mathbf{W}}_i .
\end{equation}
It can be verified that the constructed $\{\{\bar{\mathbf{W}}_k\}_{k=1}^K, \bar{\mathbf{R}}_s\}$ can meet constraints~\eqref{RW_SINR_matrix},~\eqref{constraint:power_matrix},~\eqref{constraint:semidefinite_all}, and~\eqref{constraint:rank}, while  achieving the same objective value as solution $\{\{\hat{\mathbf{W}}_k\}_{k=1}^K, \hat{\mathbf{R}}_s\}$. Thus, $\{\{\bar{\mathbf{W}}_k\}_{k=1}^K, \bar{\mathbf{R}}_s\}$ is the optimal solution of problem~\eqref{problem:CRB_WR_U_matrix}.

\subsection{Optimize $\mathbf{X}$ with Fixed $\mathbf{W}$ and $\mathbf{R}_s$}\label{subsection_III_B}

Fixing $\mathbf{W}$ and $\mathbf{R}_s$, the pinching beamforming sub-problem reduces to optimizing the deployment $\mathbf{X}$ of the PASS antennas:
\begin{subequations}\label{problem:CRB_X}
    \begin{align}        
        \min_{ \mathbf{X}} \quad &  \text{tr}(\text{CRB}(x_t, y_t)) \\
        \mathrm{s.t.} \quad & \eqref{constraint:PASS}, \eqref{constraint:SINR}. 
    \end{align}
\end{subequations}
The PA deployment problem~\eqref{problem:CRB_X} is non-convex due to the complex relationship between the sensing CRB/communication SINR and the PA deployment. What's more, the location of PAs is highly coupled, which makes the problem more intractable. In the following, we introduce auxiliary matrices to decompose the optimization variables and adopt the AO technique along with the penalty method and SCA to tackle this problem.

Note that the overall
channel between the PASS and user $k$ given in~\eqref{h_k} can be expressed as 
\begin{equation}
\begin{aligned}
    &\mathbf{h}_k=\left[\mathbf{f}_1^\mathrm{T}\mathbf{h}_k^1,\mathbf{f}_2^\mathrm{T}\mathbf{h}_k^2,\cdots,\mathbf{f}_N^\mathrm{T}\!\mathbf{h}_k^N\right]^\mathrm{T}=\\
    &\sum\nolimits_{m=1}^M\!\!\left[\left[\mathbf{f}_1\right]_m\!\left[\mathbf{h}_k^1\right]_m\!,\left[\mathbf{f}_2\right]_m\left[\mathbf{h}_k^2\right]_m\!,\cdots\!,\left[\mathbf{f}_N\right]_m\!\left[\mathbf{h}_k^N\right]_m\right]^\mathrm{T}\\&=\sum\nolimits_{m=1}^M\mathbf{h}_{k,m},
\end{aligned}   
\end{equation}
where $\mathbf{h}_{k,m}$ stands for the channel between the $m$-th PA on all $N$ waveguides and user $k$. Based on this, the channel between the $m$-th PA on all $N$ waveguides and all $k$ communication users can be expressed as
\begin{equation}
        \mathbf{H}_m=\left[\mathbf{h}_{1,m},\mathbf{h}_{2,m},\cdots,\mathbf{h}_{K,m}\right]\in \mathbb{C}^{N \times K}.
\end{equation}
Then, the channel between the PASS and all $k$ communication users can be expressed as
\begin{equation}
        \mathbf{H}=\left[\mathbf{h}_{1},\mathbf{h}_{2},\cdots,\mathbf{h}_{K}\right]=\sum\nolimits_{m=1}^M\mathbf{H}_m\in \mathbb{C}^{N \times K}.
\end{equation}
Define auxiliary matrices and vectors 
\begin{equation}
        \mathbf{Q}=\left[\mathbf{q}_{1},\mathbf{q}_{2},\cdots,\mathbf{q}_{K}\right]=\sum\nolimits_{m=1}^M\mathbf{Q}_m\in \mathbb{C}^{N \times K}.
\end{equation}
Problem~\eqref{problem:CRB_X} can be reformulated as 
\begin{subequations}\label{problem:CRB_X_Q}
    \begin{align}        
        \min_{ \mathbf{X},\mathbf{Q},\left\{\mathbf{Q}_m\right\}_{m=1}^M} \quad \!\!&  \text{tr}(\text{CRB}(x_t, y_t)) \\
        \mathrm{s.t.} \quad & \eqref{constraint:PASS}, \\
        \label{QH}& \mathbf{Q}_m=\mathbf{H}_m, \forall m \in \mathcal{M},\\ 
        \label{Qm}& \mathbf{Q}=\sum\nolimits_{m=1}^M\mathbf{Q}_m,\\ 
        \label{SINR_Q}& \frac{|\mathbf{q}_{k}^\mathrm{H}\mathbf{w}_{k}|^2}{\sum\nolimits_{i\neq k}|\mathbf{q}_{k}^\mathrm{H}\mathbf{w}_{i}|^2+\mathbf{q}_{k}^\mathrm{H}\mathbf{R}_{s}\mathbf{q}_{k}+\sigma_0^2}\ge\gamma_k,\forall k .
    \end{align}
\end{subequations}
Adopting the penalty method  to tackle constraints~\eqref{QH} and~\eqref{Qm}, the resulting optimization problem can be given as 
\begin{subequations}\label{problem:CRB_X_Q_penalty}
    \begin{align}        
        \min_{ \mathbf{X},\mathbf{Q},\left\{\mathbf{Q}_m\right\}_{m=1}^M} \quad \!\!&  \text{tr}(\text{CRB}(x_t, y_t))+\frac{1}{\rho}\left\|\mathbf{Q}-\sum\nolimits_{m=1}^M\mathbf{Q}_m\right\|_F^2 \\\nonumber&+\frac{1}{\rho} \sum\nolimits_{m=1}^M\left\|\mathbf{Q}_m-\mathbf{H}_m\right\|_F^2\\
        \mathrm{s.t.} \quad & \eqref{constraint:PASS},\eqref{SINR_Q} ,
    \end{align}
\end{subequations}
where $\rho$ is the penalty factor. Problem~\eqref{problem:CRB_X_Q_penalty} is still non-convex due to the non-convex CRB term in the objective function and the non-convex fractional constraint~\eqref{SINR_Q}. The optimization variable matrix $\mathbf{X}$ is highly coupled with the auxiliary matrices $\mathbf{Q}$ and $\mathbf{Q}_m$, the AO technique is adopted to iteratively solve problem~\eqref{problem:CRB_X_Q_penalty}. In each iteration, problem~\eqref{problem:CRB_X_Q_penalty} is sequentially solved with respect to $\mathbf{Q}$,  $\mathbf{Q}_m$, and $\mathbf{X}$, respectively, while keeping other variables fixed.

\subsubsection{Solving the sub-problem with respect to $\mathbf{Q}$} With fixed $\mathbf{X}$ and $\mathbf{Q}_m, \forall m$, the sub-problem with respect to $\mathbf{Q}$ can be expressed as
\begin{subequations}\label{problem:CRB_X_Q_penalty_Q}
    \begin{align}        
        \min_{ \mathbf{Q}} \quad &  \left\|\mathbf{Q}-\sum\nolimits_{m=1}^M\mathbf{Q}_m\right\|_F^2 \\
        \mathrm{s.t.} \quad & \eqref{SINR_Q}.
    \end{align}
\end{subequations}
Problem~\eqref{problem:CRB_X_Q_penalty_Q} is non-convex due to the non-convex constraint~\eqref{SINR_Q}, which can reformulated as 
\begin{equation}
     \sum\nolimits_{i\neq k}\mathbf{q}_{k}^\mathrm{H}\mathbf{W}_{i}\mathbf{q}_{k}+\mathbf{q}_{k}^\mathrm{H}\mathbf{R}_{s}\mathbf{q}_{k}-\frac{1}{\gamma_k}\mathbf{q}_{k}^\mathrm{H}\mathbf{W}_{k}\mathbf{q}_{k}+\sigma_0^2 \le 0,\forall k \in \mathcal{K}.
\end{equation}
The reformulated constraint is a difference of convex (DC) functions, where the first two terms and the constant are convex quadratic forms, while the third term, $-\frac{1}{\gamma_k}\mathbf{q}_{k}^\mathrm{H}\mathbf{W}_{k}\mathbf{q}_{k}$, is concave due to the negative coefficient.

To address this non-convexity, we adopt the SCA method, which iteratively approximates the concave part of the constraint with its first-order Taylor expansion around a feasible point, thereby converting the original non-convex constraint into a convex surrogate that can be efficiently handled at each iteration. Specifically, in the $t$-th iteration, the non-convex term is linearized using its first-order Taylor expansion around a feasible point $\mathbf{q}_{k}^{(t)}$, which yields a global upper bound due to its concavity. The resulting linear approximation is given by
\begin{equation}
\begin{aligned}
        -\frac{1}{\gamma_k}&\mathbf{q}_{k}^\mathrm{H}\mathbf{W}_{k}\mathbf{q}_{k} \le -\frac{1}{\gamma_k}\left(\mathbf{q}_{k}^{(t)}\right)^\mathrm{H}\mathbf{W}_{k}\mathbf{q}_{k}^{(t)} \\
        &-\frac{1}{\gamma_k}\mathbf{q}_{k}^\mathrm{H}\mathbf{W}_{k}\left(\mathbf{q}_{k}-\mathbf{q}_{k}^{(t)}\right)-\frac{1}{\gamma_k}\left(\mathbf{q}_{k}-\mathbf{q}_{k}^{(t)}\right)^\mathrm{H}\mathbf{W}_{k}\mathbf{q}_{k}.
\end{aligned}
\end{equation}
By substituting the above into the original constraint, we obtain the following convex surrogate constraint at iteration $t$ as~\eqref{SINR_Q_SCA} at the top of the next page.
\begin{figure*}[!t]
\normalsize
\begin{equation}\label{SINR_Q_SCA}
\begin{aligned}
     \sum\nolimits_{i\neq k}\mathbf{q}_{k}^\mathrm{H}\mathbf{W}_{i}\mathbf{q}_{k}+\mathbf{q}_{k}^\mathrm{H}\mathbf{R}_{s}\mathbf{q}_{k}-\frac{1}{\gamma_k}\mathbf{q}_{k}^\mathrm{H}\mathbf{W}_{k}\left(\mathbf{q}_{k}-\mathbf{q}_{k}^{(t)}\right)-\frac{1}{\gamma_k}\left(\mathbf{q}_{k}-\mathbf{q}_{k}^{(t)}\right)^\mathrm{H}\mathbf{W}_{k}\mathbf{q}_{k}-\frac{1}{\gamma_k}\left(\mathbf{q}_{k}^{(t)}\right)^\mathrm{H}\mathbf{W}_{k}\mathbf{q}_{k}^{(t)}+\sigma_0^2 \le 0.
\end{aligned}
\end{equation}
\hrulefill \vspace*{0pt}
\end{figure*}
In this way, the originally non-convex problem is transformed into a convex one, which can be expressed as
\begin{subequations}\label{problem:CRB_X_Q_penalty_Q_SCA}
    \begin{align}        
        \min_{ \mathbf{Q}} \quad &  \left\|\mathbf{Q}-\sum\nolimits_{m=1}^M\mathbf{Q}_m\right\|_F^2 \\
        \mathrm{s.t.} \quad & \eqref{SINR_Q_SCA}.
    \end{align}
\end{subequations}
The approximated problem in each SCA iteration can be efficiently handled using CVX~\cite{grant2014cvx}. 

\begin{algorithm}[t]
\caption{SCA algorithm for solving problem~\eqref{problem:CRB_X_Q_penalty_Q}.}\label{algorithm1}
\begin{algorithmic}[1]
\STATE {Initialize feasible auxiliary matrix $\mathbf{Q}^{(0)}=\left[\mathbf{q}_1^{(0)},\mathbf{q}_2^{(0)},\cdots,\mathbf{q}_K^{(0)}\right]$, and the convergence tolerance $\epsilon_{\ref{algorithm1}} > 0$, and set the iteration index $t=0$.}
\REPEAT
\STATE   Solve  approximation problem~\eqref{problem:CRB_X_Q_penalty_Q_SCA} at local point $\mathbf{Q}^{(t)}$. 
\STATE   Update $\mathbf{Q}^{(t+1)}$ with solutions in step $3$. Set $t=t+1$. 
\UNTIL the objective function value of problem~\eqref{problem:CRB_X_Q_penalty_Q} experiences a fractional increase smaller than~$\epsilon_{\ref{algorithm1}}$.
\end{algorithmic}
\end{algorithm}

The SCA-based algorithm proceeds by iteratively solving the convexified problem and updating the linearization point $\mathbf{q}_{k}^{(t)}$ until convergence, which is summarized in \textbf{Algorithm~\ref{algorithm1}}. The proposed SCA-based algorithm ensures that the objective value is non-increasing and converges to a stationary point of the original problem~\eqref{problem:CRB_X_Q_penalty_Q}. The computational complexity of \textbf{Algorithm~\ref{algorithm1}} mostly originates from solving problem~\eqref{problem:CRB_X_Q_penalty_Q_SCA} using the interior point technique. Updating $\mathbf{Q}$ involves solving problem~\eqref{problem:CRB_X_Q_penalty_Q_SCA}, where $NK$ variables are optimized subject to $K$ second order cone (SOC) constraints of dimension $NK$, leading to a complexity of $\mathcal{O}\left(\log(\frac{1}{\epsilon_{\ref{algorithm1}}})N^{3}K^{3.5}\right)$~\cite{boyd2004convex}.

\begin{algorithm}[t]
\caption{Element-wise algorithm for solving problem~\eqref{problem:CRB_X_Q_penalty_X}.}\label{algorithm2}
\begin{algorithmic}[1]
\STATE {Initialize feasible PA deployment variable set $\mathcal{X}^{(0)}=\left\{{x_m^n}^{(0)},\forall m,n\right\}$, and the convergence tolerance $\epsilon_{\ref{algorithm2}} > 0$, and set the iteration index $t=0$.}
\REPEAT
\FOR{$n=1$ to $N$}
\FOR{$m=1$ to $M$}
\STATE Update ${x_m^n}^{(t+1)}$ by solving problem~\eqref{problem:CRB_X_Q_penalty_X_mn}.
\ENDFOR
\ENDFOR
\STATE Obtain $\mathcal{X}^{(t+1)}$ through step $3$ to step $7$. Set $t=t+1$. 
\UNTIL the objective function value of problem~\eqref{problem:CRB_X_Q_penalty_X} experiences a fractional increase smaller than~$\epsilon_{\ref{algorithm2}}$.
\end{algorithmic}
\end{algorithm}

\subsubsection{Solving the sub-problem with respect to $\mathbf{Q}_m, \forall m$} With fixed $\mathbf{X}$ and $\mathbf{Q}$, the sub-problem with respect to $\mathbf{Q}_m, \forall m$, can be expressed as
\begin{subequations}\label{problem:CRB_X_Q_penalty_Qm}
    \begin{align}        
        \min_{ \left\{\mathbf{Q}_m\right\}_{m=1}^M } \quad &   \left\|\mathbf{Q}-\sum\nolimits_{m=1}^M\mathbf{Q}_m\right\|_F^2+\sum\nolimits_{m=1}^M\left\|\mathbf{Q}_m-\mathbf{H}_m\right\|_F^2.
    \end{align}
\end{subequations}
Problem~\eqref{problem:CRB_X_Q_penalty_Qm} is an unconstrained convex optimization problem, whose closed-form optimal solution can be obtained by setting its derivative to zero. Denote the objective function of problem~\eqref{problem:CRB_X_Q_penalty_Qm} as $f\left(\mathbf{Q}_1,\mathbf{Q}_2,\cdots,\mathbf{Q}_M\right)$. The corresponding solution is given by 
\begin{equation}
       \mathbf{Q}_m+\sum_{i=1}^M \mathbf{Q}_i-\mathbf{Q}-\mathbf{H}_m = \mathbf{0}_{N \times K}, \forall m.
\end{equation}
Based on the fact that $\sum_{m=1}^{M}{\partial f\left(\mathbf{Q}_1,\mathbf{Q}_2,\cdots,\mathbf{Q}_M\right)}/{\partial {\mathbf{Q}_m}}=\mathbf{0}_{N \times K}$, we can solve that 
\begin{equation}\label{opt_Qm}
        \mathbf{Q}_m^*=\mathbf{B}+\mathbf{H}_m, \forall m,
\end{equation}
where $\mathbf{B}=\frac{1}{M+1}\left(\mathbf{Q}-\sum_{i=m}^M\mathbf{H}_i\right)$.  For the update of $\mathbf{Q}_m$ for all $m$, the computation of the closed-form expression in~\eqref{opt_Qm} incurs a complexity of $\mathcal{O}\left(MNK\right)$. 

\begin{algorithm}[t]
\caption{Penalty algorithm for solving problem~\eqref{problem:CRB_X}.}\label{algorithm3}
\begin{algorithmic}[1]
\STATE {Initialize feasible auxiliary matrix  $\mathbf{Q}_m^{(0)}$ for all $m$. PA deployment variable set $\mathcal{X}^{(0)}$, the AO convergence tolerance $\epsilon_{\ref{algorithm3}}^1 > 0$, the penalty iteration convergence tolerance $\epsilon_{\ref{algorithm3}}^2 > 0$, the penalty factor $\rho$,  and the penalty update factor $\mu<1$. Set the iteration index  $t=0$.}
\REPEAT
\REPEAT
\STATE Given $\mathbf{Q}_m^{(t)}$ for all $m$ and $\mathcal{X}^{(t)}$, update $\mathbf{Q}^{(t+1)}$ by solving problem~\eqref{problem:CRB_X_Q_penalty_Q} using \textbf{Algorithm~\ref{algorithm1}}, initialized with the previous  feasible auxiliary matrix $\mathbf{Q}^{(t)}$.
\STATE   Given $\mathbf{Q}^{(t+1)}$ and $\mathcal{X}^{(t)}$, update $\mathbf{Q}_m^{(t+1)}$ for all $m$ with~\eqref{opt_Qm}. 
\STATE Given $\mathbf{Q}^{(t+1)}$ and $\mathbf{Q}_m^{(t+1)}$ for all $m$ , update $\mathcal{X}^{(t+1)}$ by solving problem~\eqref{problem:CRB_X_Q_penalty_X_mn} using \textbf{Algorithm~\ref{algorithm2}}, initialized with the previous  PA deployment $\mathcal{X}^{(t)}$. 
\STATE Set the iteration index $t=t+1$.
\UNTIL the objective function value of problem~\eqref{problem:CRB_X_Q_penalty_Q} experiences a fractional increase smaller than~$\epsilon_{\ref{algorithm3}}^2$.
\STATE Reduce the penalty factor with $\rho = \mu\rho$. 
\UNTIL the constraint violation is below threshold $\epsilon_{\ref{algorithm3}}^1$. 
\end{algorithmic}
\end{algorithm}

\subsubsection{Solving  the sub-problem with respect to $\mathbf{X}$} With fixed $\mathbf{Q}$ and $\mathbf{Q}_m, \forall m$, the sub-problem with respect to $\mathbf{X}$ can be expressed as
\begin{subequations}\label{problem:CRB_X_Q_penalty_X}
    \begin{align}        
        \min_{ \mathbf{X} } \quad &  \text{tr}(\text{CRB}(x_t, y_t))+\frac{1}{\rho} \sum\nolimits_{m=1}^M\left\|\mathbf{Q}_m-\mathbf{H}_m\right\|_F^2\\
        \mathrm{s.t.} \quad & \eqref{constraint:PASS}.
    \end{align}
\end{subequations}
Optimization variables in problem~\eqref{problem:CRB_X_Q_penalty_X} are highly coupled in the CRB term of the objective function and constraint~\eqref{constraint:PASS}. To handle this challenging problem, an element-wise optimization is adopted to iteratively optimize the location of each PA while keeping other PAs fixed. The optimization problem with respect to the location parameters of the $m$-th PA on the $n$-th waveguide of PASS $x_m^n$ can be expressed as
\begin{subequations}\label{problem:CRB_X_Q_penalty_X_mn}
    \begin{align}        
        \min_{ x_m^n } \quad &  \text{tr}(\text{CRB}(x_t, y_t))+\frac{1}{\rho} \sum\nolimits_{k=1}^K\left|\left[\mathbf{Q}_m\right]_{nk}-\left[\mathbf{H}_m\right]_{nk}\right|^2\\
        \mathrm{s.t.} \quad
        & 0 \le  x_m^n   \le x_{m+1}^n - \delta, \text{if}\ m=1,\\ 
        & x_{m-1}^n + \delta \le  x_m^n   , \text{if}\ 2 \le m \le M-1,
        \\
        & x_{m-1}^n + \delta \le  x_m^n   \le L, \text{if}\ m=M.
    \end{align}
\end{subequations}
Problem~\eqref{problem:CRB_X_Q_penalty_X_mn} is a one dimensional optimization problem which can be solved with one dimensional search. Specifically, the continuous variable $x_m^n$ is uniformly discretized over its feasible domain into $Q$ fine grids. Let $\mathcal{X}_m^n $ denote the set of discretized feasible candidates. Then, the solution of the one dimensional search is 
\begin{equation}
     {x_m^n}^*\!\!=\!\mathop{\arg\min}\limits_{ x_m^n\in \mathcal{X}_m^n }  \text{tr}(\text{CRB}(x_t, y_t))+\frac{1}{\rho} \sum\nolimits_{k=1}^K\!\left|\left[\mathbf{Q}_m\right]_{nk}-\left[\mathbf{H}_m\right]_{nk}\right|^2\!\!. 
\end{equation}
The element-wise algorithm proceeds by iteratively solving the single PA placement problem and updating the PA deployment setting until convergence, which is summarized in \textbf{Algorithm~\ref{algorithm2}}. Updating $\mathbf{X}$
requires running \textbf{Algorithm~\ref{algorithm2}}, which has a complexity of $\mathcal{O}\left(\log(\frac{1}{\epsilon_{\ref{algorithm2}}})QMNK\right)$.

\subsubsection{The penalty algorithm for solving problem~\eqref{problem:CRB_X}} 

The penalty algorithm for solving problem~\eqref{problem:CRB_X} is summarized in \textbf{Algorithm~\ref{algorithm3}}. With the given penalty factor,
the reformulated penalty PA deployment problem~\eqref{problem:CRB_X} is tackled by sequentially solving the sub-problems with respect to variables $\mathbf{Q}$, $\left\{\mathbf{Q}_m\right\}_{m=1}^M$, and $\mathbf{X}$ in each iteration. The proposed penalty algorithm ensures that the objective value is non-increasing and converges to a stationary point of the original problem~\eqref{problem:CRB_X}. The computational complexity of \textbf{Algorithm~\ref{algorithm3}} mostly originates from solving problem~\eqref{problem:CRB_X_Q_penalty_Q} using \textbf{Algorithm~\ref{algorithm1}} and can be expressed as $\mathcal{O}\left(AN^{3}K^{3.5}+BMNK\right)$, where $A=\log\left(\frac{1}{\epsilon_{\ref{algorithm3}}^1}\right)\log\left(\frac{1}{\epsilon_{\ref{algorithm3}}^2}\right)$, $B=A\log\left(\frac{1}{\epsilon_{\ref{algorithm1}}}\right)$, $C=A\left(\log\left(\frac{1}{\epsilon_{\ref{algorithm2}}}\right)Q+1\right)$.

\begin{algorithm}[t]
\caption{Overall AO algorithm for solving problem~\eqref{problem:CRB}.}\label{algorithm4}
\begin{algorithmic}[1]
\STATE {Initialize  PA deployment variable set $\mathcal{X}^{(0)}$, and the convergence tolerance $\epsilon_{\ref{algorithm4}} > 0$. Set the iteration index  $t=0$.}
\REPEAT
\STATE Given $\mathcal{X}^{(t)}$, update $\mathbf{W}^{(t+1)}$ and $\mathbf{R}_s^{(t+1)}$ by solving problem~\eqref{problem:CRB_X_Q_penalty_Q}. 
\STATE Given $\mathbf{W}^{(t+1)}$ and $\mathbf{R}_s^{(t+1)}$ , update $\mathcal{X}^{(t+1)}$ by solving problem~\eqref{problem:CRB_X_Q_penalty_X_mn} using \textbf{Algorithm~\ref{algorithm3}}. 
\STATE Set the iteration index $t=t+1$.
\UNTIL the objective function value of problem~\eqref{problem:CRB} experiences a fractional increase smaller than~$\epsilon_{\ref{algorithm4}}$.
\end{algorithmic}
\end{algorithm}

\subsection{The overall AO algorithm for solving problem~\eqref{problem:CRB}} \label{subsection_III_C}

To tackle the proposed joint PA deployment and beamforming problem~\eqref{problem:CRB}, we employ the AO framework. Specifically, the beamforming sub-problem~\eqref{problem:CRB_WR} and the PA deployment sub-problem~\eqref{problem:CRB_X_Q_penalty_Q} are addressed in Section~\ref{subsection_III_A} and Section~\ref{subsection_III_B}, respectively. The AO algorithm proceeds by iteratively solving the divided sub-problems and updating the beamforming and PA deployment setting until convergence, which is summarized in \textbf{Algorithm~\ref{algorithm4}}. The proposed AO algorithm ensures that the objective value is non-increasing and converges to a stationary point of the original problem~\eqref{problem:CRB}. The computational complexity of \textbf{Algorithm~\ref{algorithm4}}  originates from solving problem~\eqref{problem:CRB_WR} and solving problem~\eqref{problem:CRB_X} using \textbf{Algorithm~\ref{algorithm3}}, which can be expressed as $\mathcal{O}\left(\log(\frac{1}{\epsilon_{\ref{algorithm4}}})\left(\left(K+1\right)^{4.5}N^{4.5}+AN^{3}K^{3.5}+BMNK\right)\right)$.

\section{Simulation Results} \label{sec:results}
In this section, the simulation results of the proposed PASS assisted ISAC system are provided to verify the advantages of PASS and the effectiveness of the proposed algorithms.
\subsection{Simulation Setup}
\begin{figure} [htbp]
\centering
\includegraphics[width=0.5\textwidth]{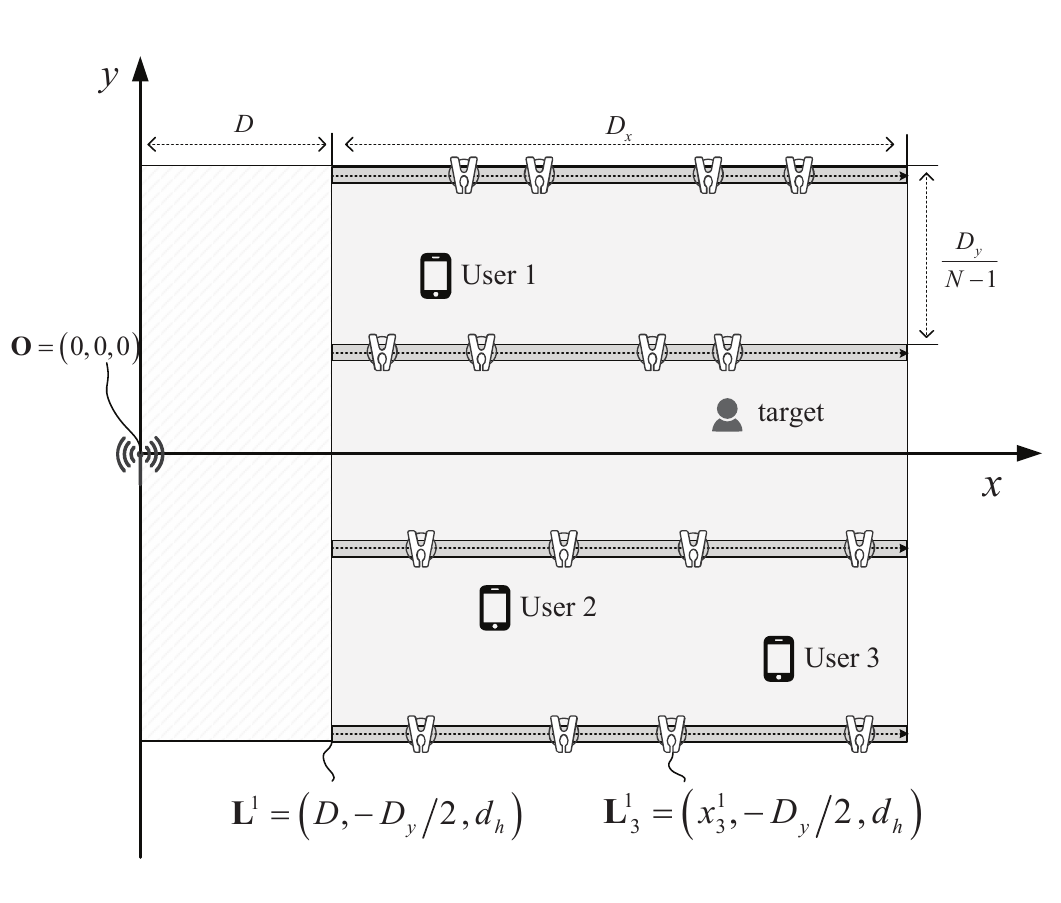}
 \caption{Simulation setup for the PASS assisted ISAC system.
  }
 \label{system_setup}
\end{figure}

The considered PASS assisted ISAC system is illustrated in Fig.~\ref{system_setup}, where the BS is located at the origin of the Cartesian coordinate system. The service area is defined on the $X$–$Y$ plane, within which all communication users and the sensing target are uniformly distributed.

{\color{blue}In particular, the PASS contains $N=4$ waveguides, each equipped with $M=4$ PAs, and the minimum separation between adjacent PAs is set to $\delta=\lambda_c/2$. }The effective refractive index of each waveguide is $n_e=1.4$, and the attachable length of the waveguides is $L=15$ m. The BS is located at a distance of $D=5$ m from the service area with a PASS height of $d_h=1$ m. The service area spans $D_x \times D_y = 15 \text{m} \times 15 \text{m}$. The BS is equipped with $M_r=8$ receive antennas, and serves $K=3$ single-antenna communication users.

The length of the coherent processing interval is $T=256$. The carrier frequency is set as $f_c=28$ GHz. The communication noise power at each user terminal is $\sigma_0^2=-90$ dBm, while the BS receiver noise power is set to $\sigma_s^2=-90$ dBm. The BS has a total transmit power budget of $P=30$ dBm, with a PA power allocation coefficient of $\rho=1/\sqrt{M}$. Each communication user is required to achieve a target SINR of $\gamma_k=\gamma=6$ dB, $\forall k \in \mathcal{K}$. {\color{blue}Unless otherwise specified, the parameters adopted in the subsequent simulations are consistent with those mentioned above.}

\subsection{Baseline Schemes}
In order to evaluate the performance of the proposed PASS assisted ISAC system, several baseline schemes are considered for comparison. These baselines include conventional MIMO architectures as well as PASS assisted schemes with different PA deployment strategies. The descriptions of these schemes are provided as follows:

\begin{enumerate}
\item {\textbf{Traditional MIMO ISAC Scheme:} In this baseline scheme,  the BS is equipped with a traditional MIMO full digital transmitter whose antenna number is set as~$N$.}

\item {\textbf{Massive MIMO ISAC Scheme~\cite{10579914}:} In this baseline scheme,  the BS is equipped with a massive MIMO transmitter. To guarantee fair comparison, the transmitter of the BS adopted a partially connected hybrid beamforming architecture, where the number of radio frequency (RF) chains is set as $N$ while the number of antennas connected to each RF chain is $M$.}

\item {\textbf{PASS assisted ISAC Scheme with Uniformly Deployed PAs:} In this baseline scheme, the PAs are uniformly placed along each waveguide. This scheme can simultaneously exploit the benefits of PASS while reducing the deployment complexity.}

\item \textbf{PASS assisted ISAC Scheme with discrete PA design:} In this baseline scheme, the PAs can only be attached to~$Z+1$ predefined discrete positions along each waveguide. Under this setting, the feasible set of the $x$-axis coordinates of the PAs can be expressed as 
 \begin{equation}\label{X_D}
\begin{aligned}
\hat{\mathcal{X}}=\left\{ {\begin{array}{*{20}{c}}
{x_m^n,}\!\!\!&\vline& \!\!\!{ x_m^n \in \left\{ D, D+\frac{L}{Z}, \cdots, D+L \right\} \!\!}\\
{ \forall n, m}\!\!\!&\vline& \!\!\!{x_m^n - x_{m-1}^n \ge \delta}
\end{array}} \right\}.
\end{aligned}
\end{equation}
The optimization problem corresponding to this baseline scheme can also be addressed using the algorithms proposed in this work, since the one-dimensional search method can be directly applied to the discrete PA deployment setting. 
\end{enumerate}

In the simulation figures, legend ``Continuous-PASS'' denotes the proposed PASS assisted ISAC system where the positions of the PAs can be continuously adjusted. Legend ``Discrete-PASS-$Z$'' denotes the proposed PASS assisted ISAC system where the PAs can only be attached to several predefined discrete positions, with $Z$ representing the granularity of available PA positions on each waveguide. Legend ``Uniform-PASS'' denotes the proposed PASS assisted ISAC system where the PAs are uniformly distributed along each waveguide. Legend ``mMIMO'' denotes the massive MIMO enabled ISAC system, where the BS transmitter adopts a partially connected hybrid beamforming architecture with $N$ RF chains, each connected to $M$ antennas. Legend ``MIMO'' denotes the conventional MIMO enabled ISAC system, where the BS transmitter adopts a fully digital beamforming architecture with $N$ RF chains.
\begin{figure} [t]
\centering
\includegraphics[width=0.5\textwidth]{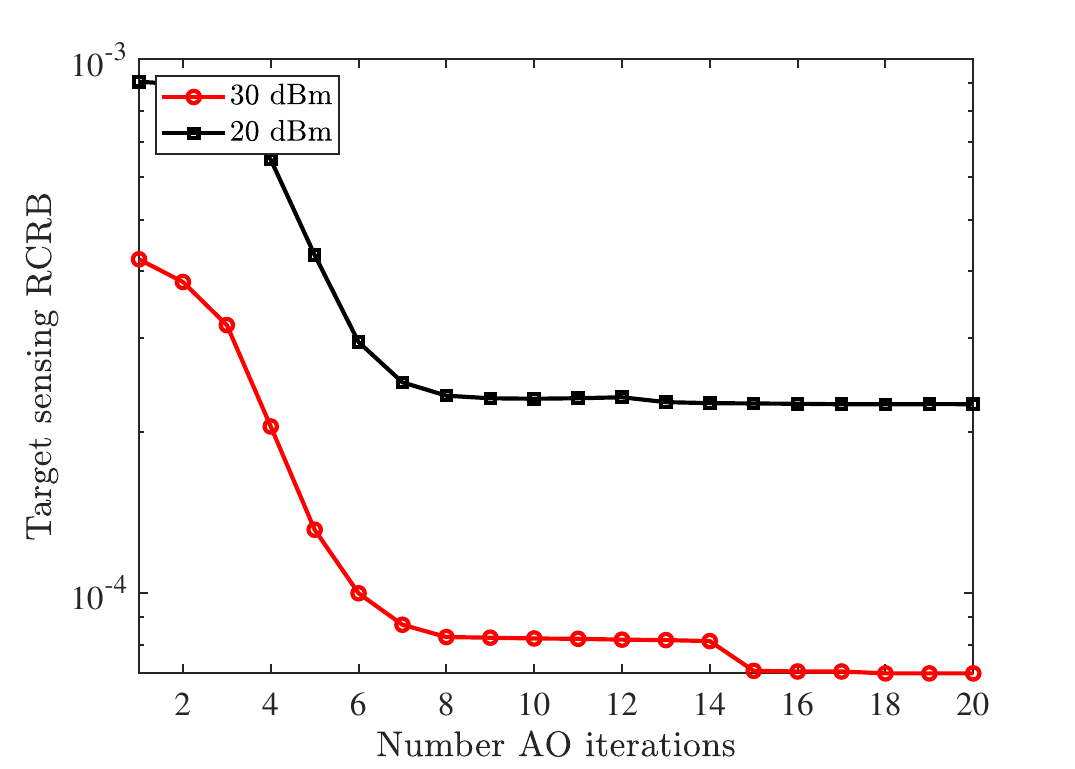}
 \caption{Convergence behavior of proposed \textbf{Algorithm~\ref{algorithm4}}.}
 \label{convergence}
\end{figure}
\begin{figure} [t]
\centering
\includegraphics[width=0.5\textwidth]{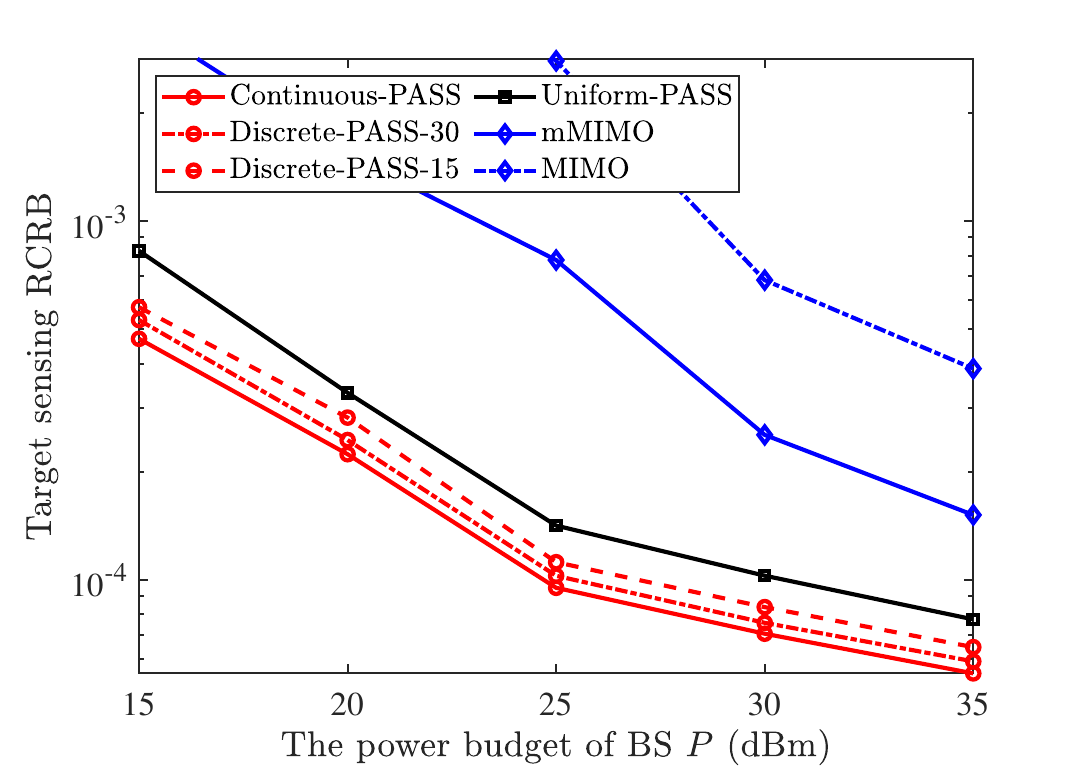}
 \caption{RCRB versus the power budget of the BS $P$.}
 \label{Power_P}
\end{figure}

\subsection{Convergence Behavior of Proposed Algorithm.}

The convergence behavior of the proposed \textbf{Algorithm~\ref{algorithm4}} is illustrated in Fig.~\ref{convergence}, where the root CRB (RCRB) is plotted against the number of AO iterations for two different base station power budgets, i.e., $P=20$ dBm and $P=30$ dBm. It can be observed that the proposed algorithm converges rapidly, achieving stable RCRB values within approximately $20$ iterations for both power settings. Fig.~\ref{convergence} demonstrates the efficiency of the proposed algorithm in jointly optimizing sensing, communication functionalities and PAs deployment. Besides, the convergence behavior of the massive MIMO ISAC benchmark scheme is also examined in Fig.~\ref{convergence}.

\subsection{Root CRB Versus the Power Budget of the BS}

Fig.~\ref{Power_P} illustrates the variation of the RCRB with respect to the power budget of the BS $P$. As expected, all baseline schemes exhibit improved sensing accuracy when more transmission power is available. The reason is that a higher transmit power leads to a stronger echo signal given the same communication QoS requirement, thereby  tightening the CRB. 

A comparison among different schemes reveals that PASS assisted ISAC consistently outperforms MIMO-ISAC across the considered power budgets. In particular, the PASS assisted ISAC with continuously deployed PAs achieves the best performance, while the PASS assisted ISAC with discretely deployed PAs with $30$ quantization levels closely approaches it. As the quantization resolution decreases (e.g., to $15$ levels), the performance gap becomes more evident. This degradation stems from the limited pinching beamforming degrees of freedom caused by discrete deployment. Nevertheless, increasing the quantization resolution effectively mitigates this performance loss. 

The uniform PASS scheme, although inferior to optimized PASS schemes, still significantly outperforms both massive MIMO-ISAC and conventional MIMO-ISAC. This is because, despite its naive beamforming design, the PASS architecture inherently places antennas closer to the communication users and sensing targets. Among the MIMO baselines, massive MIMO-ISAC achieves better sensing accuracy than traditional MIMO-ISAC, owing to its larger number of antennas, while the latter yields the poorest performance overall.

\begin{figure} [t]
\centering
\includegraphics[width=0.5\textwidth]{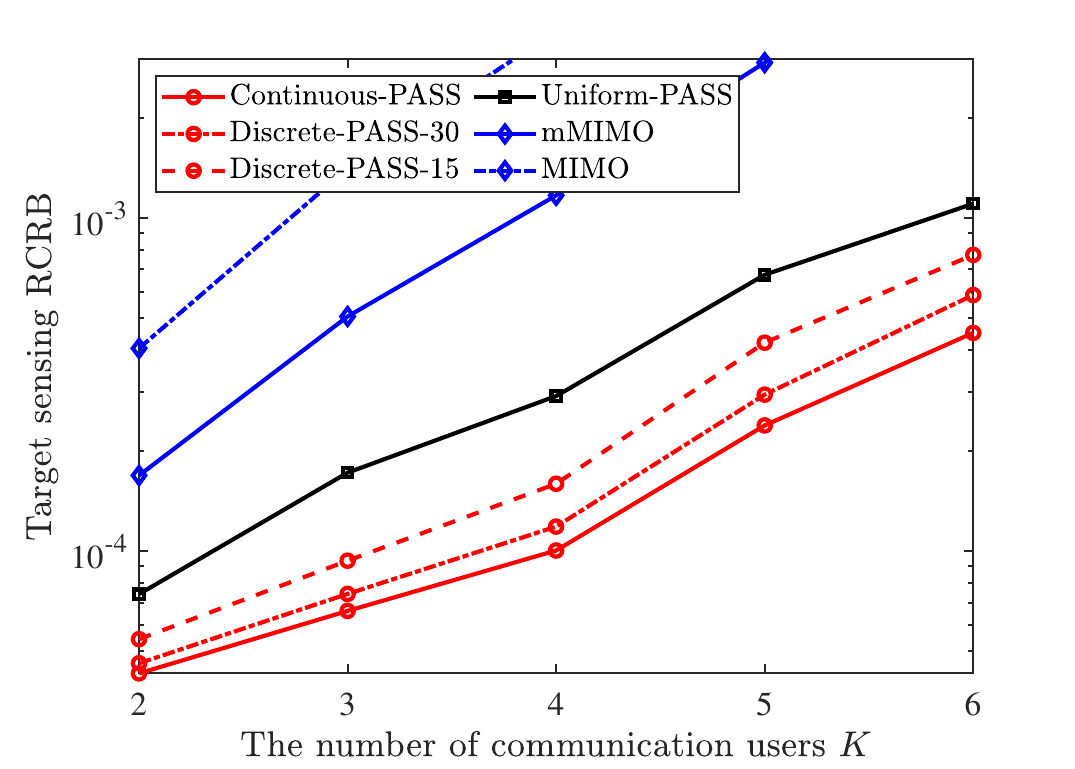}
 \caption{RCRB versus the number of users $K$.}
 \label{User_K}
\end{figure}

\begin{figure} [t]
\centering
\includegraphics[width=0.5\textwidth]{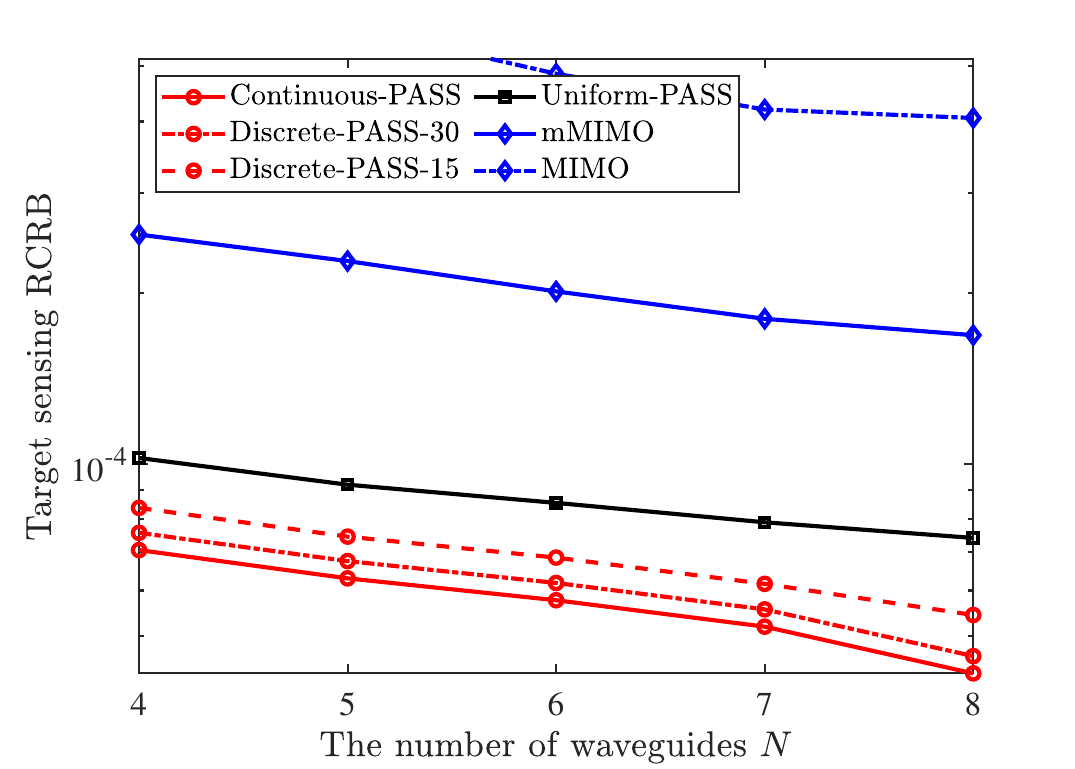}
 \caption{RCRB versus the number of waveguide $N$.}
 \label{Waveguide_N}
\end{figure}

\begin{figure} [t]
\centering
\includegraphics[width=0.5\textwidth]{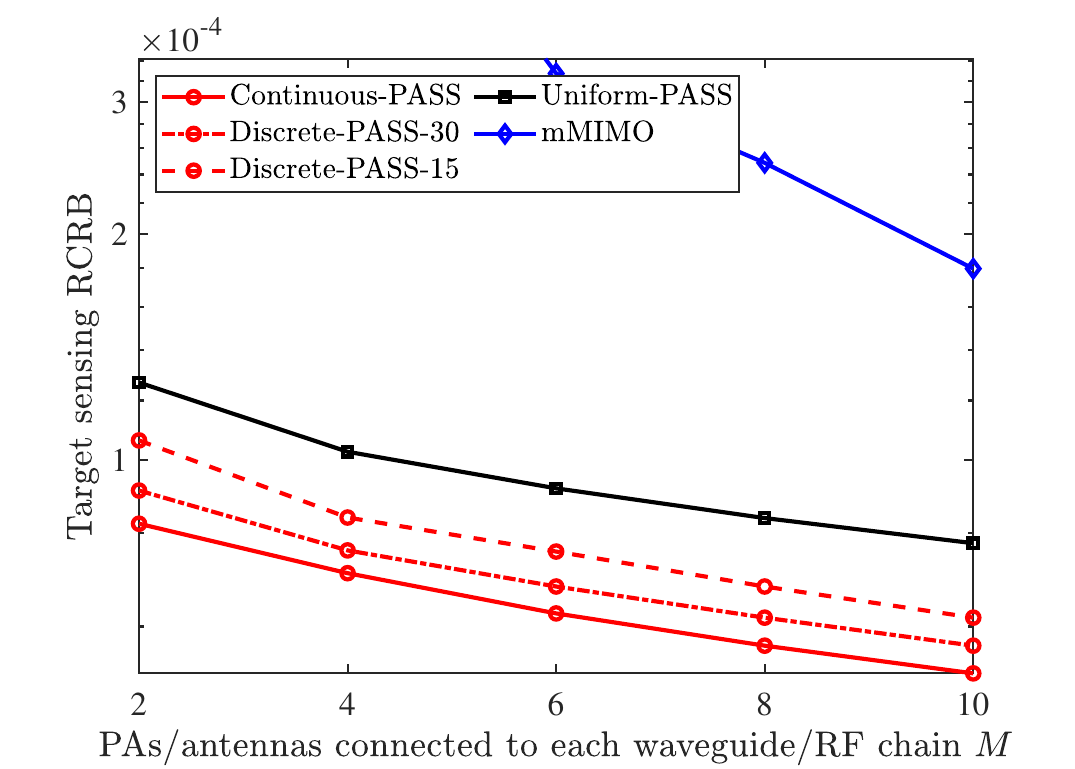}
 \caption{RCRB versus the number of PAs per waveguide~$M$.}
 \label{PA_M}
\end{figure}

\begin{figure} [t]
\centering
\includegraphics[width=0.5\textwidth]{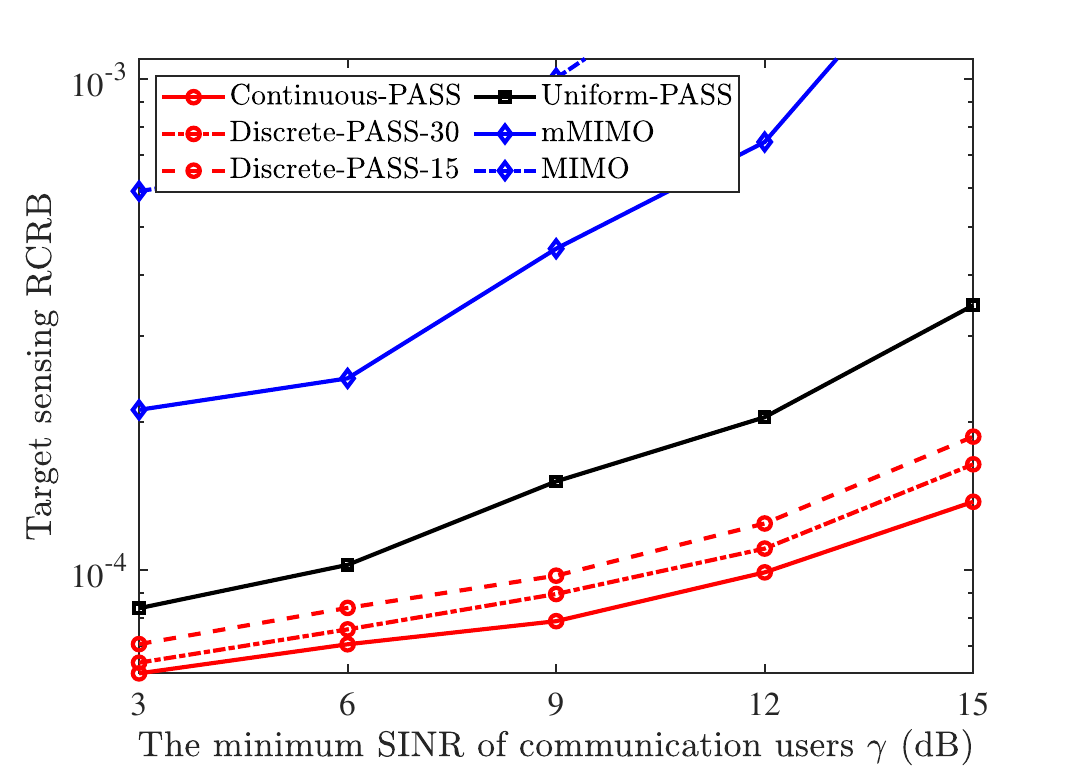}
 \caption{RCRB versus the minimum communication SINR $\gamma$.}
 \label{SINR_Gamma}
\end{figure}

\subsection{Root CRB Versus the Number of Users}

Fig.~\ref{User_K} demonstrates the RCRB versus the number of communication users $K$. The number of waveguides in PASS assisted ISAC and the number of RF chains in MIMO-ISAC schemes are set to $N=8$. The trends among different schemes are consistent with those observed in Fig.\ref{Power_P}.

Besides, as the number of communication users $K$ increases, the system RCRB rises, indicating a degradation in sensing accuracy. The underlying reason lies in the resource allocation trade-off: when more communication users are present, the limited BS resources must be more intensively allocated to satisfy the SINR requirements of communication users, thereby reducing the resources available for sensing. Additionally, as indicated by \eqref{CommunicationSINR}, the sensing signals are treated as interference for communication users. Hence, increasing the number of users imposes more stringent constraints on the sensing signal design, further deteriorating the sensing accuracy. Nevertheless, PASS assisted ISAC schemes exhibit a stronger resilience to this performance degradation compared to MIMO-ISAC schemes. This advantage stems from the capability of the PASS to reshape and enhance the wireless propagation environment through pinching beamforming, thereby providing additional degrees of freedom for joint communication and sensing design.

\subsection{Root CRB Versus the Number of Waveguides}

Fig.~\ref{Waveguide_N} demonstrates the RCRB versus the number of waveguides $N$. The trends among different schemes are consistent with those observed in previous figures. It can be observed that, across all considered schemes, the RCRB decreases as the number of waveguides increases, indicating an improvement in sensing accuracy. This improvement is primarily due to the higher number of RF chains available at the BS with more waveguides, which provides greater spatial degrees of freedom. As a result, the system can perform more flexible beamforming for both communication and sensing, thereby enhancing the overall sensing performance.

\subsection{Root CRB Versus the Number of PAs on each waveguide}

Fig.~\ref{PA_M} demonstrates the RCRB versus the number of PAs on each waveguide $M$. The trends among different schemes are consistent with those observed in previous figures. Similar to the previous observations, increasing $M$ leads to a reduction in the RCRB, reflecting enhanced sensing accuracy. This improvement can be attributed to the more precise pinching beamforming enabled by a higher number of PAs per waveguide. Specifically, having more PAs allows the system to perform communication beamforming more efficiently, thereby reserving more power for sensing. At the same time, finer pinching beamforming can generate sensing beams that are better matched to the target, further enhancing the sensing performance.

\subsection{Root CRB Versus the Minimum Communication SINR}

Fig.~\ref{SINR_Gamma} illustrates the relationship between the RCRB and the minimum communication SINR $\gamma$ of the users. Overall, the observed trends among different schemes are consistent with those reported in previous figures. Specifically, as the minimum communication SINR requirement increases, the system RCRB also increases, indicating a degradation in sensing accuracy.  This phenomenon can be attributed to the finite power and communication QoS constraints at the BS. Higher SINR requirements necessitate allocating more power and resources to the communication tasks to satisfy stringent  QoS  demands. Consequently, fewer resources remain available for sensing, leading to a higher RCRB and thus lower sensing precision. 

Moreover, while increasing the minimum communication SINR generally reduces system sensing accuracy, the impact of communication QoS constraints on the RCRB is more pronounced in traditional MIMO-ISAC systems than in PASS assisted ISAC systems. The underlying reason is that in PASS assisted ISAC systems, the  PAs  is located closer to the target, which allows the system to meet communication QoS requirements efficiently with lower transmit power. As a result, the sensing performance in PASS assisted ISAC is less sensitive to stringent communication requirements compared to conventional MIMO-ISAC schemes.

\section{Conclusions}\label{sec:conclusions}
In this work, a PASS assisted ISAC system was investigated, where a sensing CRB minimization problem was formulated under communication QoS constraints to explore the fundamental sensing accuracy limits. The system design was decomposed into a digital beamforming sub-problem at the BS baseband and a pinching beamforming sub-problem on the waveguides. To address the overall non-convex optimization problem, an AO based algorithm was developed, which iteratively solves the two sub-problems using the SDR method, the SCA method, the penalty method, and element-wise optimization techniques. Simulation results validated the effectiveness of the proposed framework, demonstrating its superior performance over benchmark schemes and highlighting the significant advantages introduced by pinching beamforming.

\section*{Appendix~A\\Derivation of the FIM for PASS assisted ISAC} \label{Appendix:A}
\renewcommand{\theequation}{A.\arabic{equation}}
\setcounter{equation}{0}
Reformulate the received PASS signal over a coherent processing interval including $T$ time slots as vectorized signal
\begin{equation}
         \text{vec}\left(\mathbf{Y}_s\right)=\text{vec}\left(\mathbf{G}\mathbf{S}\right)+\text{vec}\left(\mathbf{N}_s\right)\sim\mathcal{CN}\left(\mathbf{u}, \mathbf{R}_n\right),
 \end{equation} 
where $\mathbf{u}=\text{vec}\left(\mathbf{G}\mathbf{S}\right)$, $\mathbf{R}_n=\sigma_s^2\mathbf{I}_{M_rT}$. The element on the $i$-th row and the $j$-th column of $\mathbf{J}$ can be expressed as~\cite{kay1993fundamentals}
\begin{equation}\label{FIM_calculate}
\begin{aligned}
        \left[\mathbf{J}\right]_{ij}=\frac{2}{\sigma_s^2}\Re\left(\frac{\partial \mathbf{u}^\mathrm{H}}{\partial \eta_i} \frac{\partial \mathbf{u}}{\partial \eta_j}\right), \forall i,j \in \{1,2,3,4\},
\end{aligned}
\end{equation}
where $\eta_i$ is the $i$-th element of $\boldsymbol{\eta}$. By defining $\mathbf{A}=\mathbf{a}\left(\theta\right)\mathbf{h}_t^\mathrm{H}=\mathbf{a}\left(\theta\left({x_t},{y_t}\right)\right)\mathbf{h}_t^\mathrm{H}\left(x_t,y_t\right)$ and $\theta\left({x_t},{y_t}\right)=\arctan\left({y_t}/{x_t}\right)$, it follows that
\begin{equation}
\begin{aligned}
        &\frac{\partial \mathbf{u}}{\partial \boldsymbol{\eta}}=\left[\frac{\partial \mathbf{u}}{\partial x_t},\frac{\partial \mathbf{u}}{\partial y_t},\frac{\partial \mathbf{u}}{\partial \Re\left(\beta\right)},\frac{\partial \mathbf{u}}{\partial \Im\left(\beta\right)}\right]=\\&\left[\beta\text{vec}(\dot{\mathbf{A}}_{x_t}\mathbf{S})\!,\beta\text{vec}(\dot{\mathbf{A}}_{y_t}\mathbf{S})\!,\text{vec}(\mathbf{A}\mathbf{S})\!,j\text{vec}(\mathbf{A}\mathbf{S})\right],
\end{aligned}
\end{equation}
where
\begin{equation}
\begin{aligned}
        \dot{\mathbf{A}}_{x_t}=&\frac{\partial \mathbf{A}}{\partial x_t}=\frac{\partial \mathbf{a}\left(\arctan\left({y_t}/{x_t}\right)\right)}{\partial x_t}\mathbf{h}_t^\mathrm{H}\left(x_t,y_t\right)\\&+\mathbf{a}\left(\arctan\left({y_t}/{x_t}\right)\right)\frac{\partial \mathbf{h}_t^\mathrm{H}\left(x_t,y_t\right)}{\partial x_t},
\end{aligned}
\end{equation}
\begin{equation}
\begin{aligned}
        \dot{\mathbf{A}}_{y_t}=&\frac{\partial \mathbf{A}}{\partial y_t}=\frac{\partial \mathbf{a}\left(\arctan\left({y_t}/{x_t}\right)\right)}{\partial y_t}\mathbf{h}_t^\mathrm{H}\left(x_t,y_t\right)\\&+\mathbf{a}\left(\arctan\left({y_t}/{x_t}\right)\right)\frac{\partial \mathbf{h}_t^\mathrm{H}\left(x_t,y_t\right)}{\partial y_t},
\end{aligned}
\end{equation}
with
\begin{equation}
\begin{aligned}
        &\frac{\partial \mathbf{a}\left(\arctan\left({y_t}/{x_t}\right)\right)}{\partial x_t}=\frac{\partial \mathbf{a}\left(\theta\right)}{\partial \theta}\frac{\partial \arctan\left({y_t}/{x_t}\right)}{\partial x_t}=\\&=j\kappa_c\frac{x_ty_t}{\left(x_t^2+y_t^2\right)^{\frac{3}{2}}}\cdot\mathbf{D}\cdot\mathbf{a}\left(\arctan\left({y_t}/{x_t}\right)\right),
\end{aligned}
\end{equation}
\begin{equation}
\begin{aligned}
        &\frac{\partial \mathbf{a}\left(\arctan\left({y_t}/{x_t}\right)\right)}{\partial y_t}=\frac{\partial \mathbf{a}\left(\theta\right)}{\partial \theta}\frac{\partial\arctan\left({y_t}/{x_t}\right)}{\partial y_t}=\\&=-j\kappa_c\frac{x_t^2}{\left(x_t^2+y_t^2\right)^{\frac{3}{2}}}\cdot\mathbf{D}\cdot\mathbf{a}\left(\arctan\left({y_t}/{x_t}\right)\right),
\end{aligned}       
\end{equation}

\begin{equation}
        \mathbf{D}=\text{diag}\left(0,1,\cdots,M_r-1\right),
\end{equation}
\begin{equation}
\begin{aligned}
        &\frac{\partial \mathbf{h}_t\left(x_t,y_t\right)}{\partial x_t}=\mathbf{F}\frac{\partial \hat{\mathbf{h}}_t\left(x_t,y_t\right)}{\partial x_t}=\mathbf{F}\times\\ 
        &\!\left[\frac{\partial {\mathbf{h}}_t^{1}\left(x_t,y_t\right)^\mathrm{H}}{\partial x_t},\frac{\partial {\mathbf{h}}_t^2\left(x_t,y_t\right)^\mathrm{H}}{\partial x_t},\cdots,\frac{\partial {\mathbf{h}}_t^N\left(x_t,y_t\right)^\mathrm{H}}{\partial x_t}\right]^\mathrm{H},
\end{aligned}        
\end{equation}
\begin{equation}
\begin{aligned}
        &\frac{\partial \mathbf{h}_t\left(x_t,y_t\right)}{\partial y_t}=\mathbf{F}\frac{\partial \hat{\mathbf{h}}_t\left(x_t,y_t\right)}{\partial y_t}=\mathbf{F}\times\\ 
        &\!\left[\frac{\partial {\mathbf{h}}_t^{1}\left(x_t,y_t\right)^\mathrm{H}}{\partial y_t},\frac{\partial {\mathbf{h}}_t^2\left(x_t,y_t\right)^\mathrm{H}}{\partial y_t},\cdots,\frac{\partial {\mathbf{h}}_t^N\left(x_t,y_t\right)^\mathrm{H}}{\partial y_t}\right]^\mathrm{H},
\end{aligned}        
\end{equation}

\begin{equation}
\begin{aligned}
&\frac{\partial \left[{\mathbf{h}}_t^n\left(x_t,y_t\right)\right]_m}{\partial x_t}\\
&= - \frac{(x_t - x_m^n) \, e^{-j \kappa_c r_{t,m}^n}}{2\kappa_c} 
\left( \frac{1}{\left(r_{t,m}^n\right)^3} + \frac{j \kappa_c}{\left(r_{t,m}^n\right)^2} \right),
\end{aligned} 
\end{equation}

\begin{equation}
\begin{aligned}
&\frac{\partial \left[{\mathbf{h}}_t^n\left(x_t,y_t\right)\right]_m}{\partial y_t}\\
&= - \frac{(y_t - y^n) \, e^{-j \kappa_c r_{t,m}^n}}{2\kappa_c} 
\left( \frac{1}{\left(r_{t,m}^n\right)^3} + \frac{j \kappa_c}{\left(r_{t,m}^n\right)^2} \right).
\end{aligned} 
\end{equation}
The matrix $\mathbf{J}_{11}$ is
\begin{equation}\label{eqn:FIM_J11}
        \mathbf{J}_{11}=\begin{bmatrix}
         {J}_{x_tx_t} & {J}_{x_ty_t} \\
         {J}_{x_ty_t} & {J}_{y_ty_t} 
    \end{bmatrix}.
\end{equation}
Based on~\eqref{FIM_calculate}, the element ${J}_{i_tj_t}$, $\forall i,j\in\left\{x,y\right\}$, can be expressed as 
\begin{equation}\label{eqn:FIM_J11_ele}
      {J}_{i_tj_t}=\frac{2\left|\beta\right|^2T}{\sigma_s^2}\Re\left(\text{tr}\left(\dot{\mathbf{A}}_{j_t}\mathbf{R}\dot{\mathbf{A}}_{i_t}^\mathrm{H}\right)\right),
\end{equation} 
where we use the fact that $\text{tr}\left(\mathbf{A}\mathbf{B}\mathbf{C}\right)=\text{tr}\left(\mathbf{C}\mathbf{A}\mathbf{B}\right)$, and the sensing  covariance matrix is approximated with the  sensing signals over given $T$ time slots, i.e.,
\begin{equation}
        \mathbf{R}=\frac{1}{T}\mathbf{S}\mathbf{S^\mathrm{H}}.
\end{equation}
The matrix $\mathbf{J}_{12}$ can be expressed as 
\begin{equation}\label{eqn:FIM_J12}
\begin{aligned}
        \!\!\mathbf{J}_{12}\!=\! \frac{2 T}{\sigma_s^2} \Re \!\Big(\beta\!\left[ \mathrm{tr}(  \dot{\mathbf{A}}_{x_t} \mathbf{R}   {\mathbf{A}}^\mathrm{H} ),  \mathrm{tr}(  \dot{\mathbf{A}}_{y_t} \mathbf{R}  {\mathbf{A}}^\mathrm{H} )\right]^\mathrm{H} \!\!\left[1, j\right]\Big).
\end{aligned}
\end{equation}
The matrix $\mathbf{J}_{22}$ can be expressed as 
\begin{equation}\label{eqn:FIM_J22}
        \mathbf{J}_{22}=\frac{2T}{\sigma_s^2}\mathbf{I}_{2}\Re\left(\text{tr}\left({\mathbf{A}}\mathbf{R}{\mathbf{A}}^\mathrm{H}\right)\right).
\end{equation}

\bibliographystyle{IEEEtran}
\bibliography{myref}

\begin{thebibliography}{10}
\providecommand{\url}[1]{#1}
\csname url@samestyle\endcsname
\providecommand{\newblock}{\relax}
\providecommand{\bibinfo}[2]{#2}
\providecommand{\BIBentrySTDinterwordspacing}{\spaceskip=0pt\relax}
\providecommand{\BIBentryALTinterwordstretchfactor}{4}
\providecommand{\BIBentryALTinterwordspacing}{\spaceskip=\fontdimen2\font plus
\BIBentryALTinterwordstretchfactor\fontdimen3\font minus
  \fontdimen4\font\relax}
\providecommand{\BIBforeignlanguage}[2]{{%
\expandafter\ifx\csname l@#1\endcsname\relax
\typeout{** WARNING: IEEEtran.bst: No hyphenation pattern has been}%
\typeout{** loaded for the language `#1'. Using the pattern for}%
\typeout{** the default language instead.}%
\else
\language=\csname l@#1\endcsname
\fi
#2}}
\providecommand{\BIBdecl}{\relax}
\BIBdecl

\bibitem{li2025crb}
H.~Li, R.~Zhong, J.~Lei, P.~Zhiwen, and Y.~Liu, ``{CRB} minimization for {PASS}
  assisted {ISAC},'' \emph{arXiv preprint arXiv:2509.22181}, 2025.

\bibitem{9598915}
M.~Alsabah, M.~A. Naser, B.~M. Mahmmod, S.~H. Abdulhussain, M.~R. Eissa,
  A.~Al-Baidhani, N.~K. Noordin, S.~M. Sait, K.~A. Al-Utaibi, and F.~Hashim,
  ``6{G} wireless communications networks: A comprehensive survey,'' \emph{IEEE
  Access}, vol.~9, pp. 148\,191--148\,243, 2021.

\bibitem{6736761}
E.~G. Larsson, O.~Edfors, F.~Tufvesson, and T.~L. Marzetta, ``Massive {MIMO}
  for next generation wireless systems,'' \emph{{IEEE} Commun. Mag.}, vol.~52,
  no.~2, pp. 186--195, 2014.

\bibitem{9424177}
Y.~Liu, X.~Liu, X.~Mu, T.~Hou, J.~Xu, M.~Di~Renzo, and N.~Al-Dhahir,
  ``Reconfigurable intelligent surfaces: Principles and opportunities,''
  \emph{IEEE Commun. Surv. Tutor.}, vol.~23, no.~3, pp. 1546--1577, 2021.

\bibitem{zhu2023movable}
L.~Zhu, W.~Ma, and R.~Zhang, ``Movable antennas for wireless communication:
  Opportunities and challenges,'' \emph{{IEEE} Commun. Mag.}, vol.~62, no.~6,
  pp. 114--120, 2023.

\bibitem{wong2020fluid}
K.-K. Wong, A.~Shojaeifard, K.-F. Tong, and Y.~Zhang, ``Fluid antenna
  systems,'' \emph{{IEEE} Trans. Wireless Commun.}, vol.~20, no.~3, pp.
  1950--1962, 2020.

\bibitem{10945421}
Z.~Ding, R.~Schober, and H.~Vincent~Poor, ``Flexible-antenna systems: A
  pinching-antenna perspective,'' \emph{{IEEE} Trans. Commun.}, pp. 1--1, 2025,
  early access, doi={10.1109/TCOMM.2025.3555866}.

\bibitem{9737357}
F.~Liu, Y.~Cui, C.~Masouros, J.~Xu, T.~X. Han, Y.~C. Eldar, and S.~Buzzi,
  ``Integrated sensing and communications: Toward dual-functional wireless
  networks for 6{G} and beyond,'' \emph{{IEEE} J. Sel. Areas Commun.}, vol.~40,
  no.~6, pp. 1728--1767, 2022.

\bibitem{lu2024integrated}
S.~Lu, F.~Liu, Y.~Li, K.~Zhang, H.~Huang, J.~Zou, X.~Li, Y.~Dong, F.~Dong,
  J.~Zhu \emph{et~al.}, ``Integrated sensing and communications: Recent
  advances and ten open challenges,'' \emph{IEEE Internet Things J.}, vol.~11,
  no.~11, pp. 19\,094--19\,120, 2024.

\bibitem{liu2023snr}
R.~Liu, M.~Li, Q.~Liu, and A.~L. Swindlehurst, ``{SNR/CRB}-constrained joint
  beamforming and reflection designs for {RIS-ISAC} systems,'' \emph{{IEEE}
  Trans. Wireless Commun.}, vol.~23, no.~7, pp. 7456--7470, 2023.

\bibitem{li2025movable}
Z.~Li, J.~Ba, Z.~Su, J.~Huang, H.~Peng, W.~Chen, L.~Du, and T.~H. Luan,
  ``Movable antennas enabled {ISAC} systems: Fundamentals, opportunities, and
  future directions,'' \emph{{IEEE} Wireless Commun.}, 2025.

\bibitem{10707252}
L.~Zhou, J.~Yao, M.~Jin, T.~Wu, and K.-K. Wong, ``Fluid antenna-assisted {ISAC}
  systems,'' \emph{{IEEE} Wireless Commun. Lett.}, vol.~13, no.~12, pp.
  3533--3537, 2024.

\bibitem{10969546}
Y.~Zhang, X.~Shao, H.~Li, B.~Clerckx, and R.~Zhang, ``Full-space wireless
  sensing enabled by multi-sector intelligent surfaces,'' \emph{{IEEE} Trans.
  Wireless Commun.}, vol.~24, no.~9, pp. 7301--7316, 2025.

\bibitem{basar2021reconfigurable}
E.~Basar and I.~Yildirim, ``Reconfigurable intelligent surfaces for future
  wireless networks: A channel modeling perspective,'' \emph{{IEEE} Wireless
  Commun.}, vol.~28, no.~3, pp. 108--114, 2021.

\bibitem{wong2023fluid}
K.-K. Wong, W.~K. New, X.~Hao, K.-F. Tong, and C.-B. Chae, ``Fluid antenna
  system—part {I}: Preliminaries,'' \emph{{IEEE} Commun. Lett.}, vol.~27,
  no.~8, pp. 1919--1923, 2023.

\bibitem{yang2025pinching}
Z.~Yang, N.~Wang, Y.~Sun, Z.~Ding, R.~Schober, G.~K. Karagiannidis, V.~W. Wong,
  and O.~A. Dobre, ``Pinching antennas: Principles, applications and
  challenges,'' \emph{arXiv preprint arXiv:2501.10753}, 2025.

\bibitem{liu2025pinching}
Y.~Liu, Z.~Wang, X.~Mu, C.~Ouyang, X.~Xu, and Z.~Ding, ``Pinching-antenna
  systems ({PASS}): Architecture designs, opportunities, and outlook,''
  \emph{arXiv preprint arXiv:2501.18409}, 2025.

\bibitem{suzuki2022pinching}
H.~O.~Y. Suzuki and K.~Kawai, ``Pinching antenna: Using a dielectric waveguide
  as an antenna,'' \emph{NTT DOCOMO Technical J}, vol.~23, no.~3, pp. 5--12,
  2022.

\bibitem{zeng2025resource}
M.~Zeng, J.~Wang, O.~A. Dobre, Z.~Ding, G.~K. Karagiannidis, R.~Schober, and
  H.~V. Poor, ``Resource allocation for pinching-antenna systems:
  State-of-the-art, key techniques and open issues,'' \emph{arXiv preprint
  arXiv:2506.06156}, 2025.

\bibitem{10896748}
Y.~Xu, Z.~Ding, and G.~K. Karagiannidis, ``Rate maximization for downlink
  pinching-antenna systems,'' \emph{{IEEE} Wireless Commun. Lett.}, vol.~14,
  no.~5, pp. 1431--1435, 2025.

\bibitem{hu2025sum}
S.~Hu, R.~Zhao, Y.~Liao, D.~W.~K. Ng, and J.~Yuan, ``Sum-rate maximization for
  pinching antenna-assisted {NOMA} systems with multiple dielectric
  waveguides,'' \emph{arXiv preprint arXiv:2503.10060}, 2025.

\bibitem{bereyhi2025mimo}
A.~Bereyhi, C.~Ouyang, S.~Asaad, Z.~Ding, and H.~V. Poor, ``{MIMO-PASS}: Uplink
  and downlink transmission via {MIMO} pinching-antenna systems,'' \emph{arXiv
  preprint arXiv:2503.03117}, 2025.

\bibitem{wang2025pinching}
K.~Wang, Z.~Ding, and N.~Al-Dhahir, ``Pinching-antenna systems for physical
  layer security,'' \emph{arXiv preprint arXiv:2507.10167}, 2025.

\bibitem{jiang2025pinching}
H.~Jiang, Z.~Wang, and Y.~Liu, ``Pinching-antenna system ({PASS}) enhanced
  covert communications,'' \emph{arXiv preprint arXiv:2504.10442}, 2025.

\bibitem{mu2025pinching}
X.~Mu, G.~Zhu, and Y.~Liu, ``Pinching-antenna system ({PASS})-enabled multicast
  communications,'' \emph{arXiv preprint arXiv:2502.16624}, 2025.

\bibitem{li2025mimo}
H.~Li, Z.~Lyu, Y.~Gao, M.~Xiao, and H.~V. Poor, ``{MIMO} pinching-antenna-aided
  {SWIPT},'' \emph{arXiv preprint arXiv:2506.06754}, 2025.

\bibitem{ding2025pinching}
Z.~Ding, ``Pinching-antenna assisted {ISAC}: A {CRLB} perspective,''
  \emph{arXiv preprint arXiv:2504.05792}, 2025.

\bibitem{wang2025wireless}
Z.~Wang, C.~Ouyang, Y.~Liu, and A.~Nallanathan, ``Wireless sensing via
  pinching-antenna systems,'' \emph{arXiv preprint arXiv:2505.15430}, 2025.

\bibitem{bozanis2025cram}
D.~Bozanis, V.~K. Papanikolaou, S.~A. Tegos, and G.~K. Karagiannidis,
  ``Cram\'er-rao bounds for integrated sensing and communications in
  pinching-antenna systems,'' \emph{arXiv preprint arXiv:2505.01333}, 2025.

\bibitem{qin2025joint}
Y.~Qin, Y.~Fu, and H.~Zhang, ``Joint antenna position and transmit power
  optimization for pinching antenna-assisted {ISAC} systems,'' \emph{arXiv
  preprint arXiv:2503.12872}, 2025.

\bibitem{khalili2025pinching}
A.~Khalili, B.~Kaziu, V.~K. Papanikolaou, and R.~Schober, ``Pinching
  antenna-enabled {ISAC} systems: Exploiting look-angle dependence of {RCS} for
  target diversity,'' \emph{arXiv preprint arXiv:2505.01777}, 2025.

\bibitem{ouyang2025rate}
C.~Ouyang, Z.~Wang, Y.~Liu, and Z.~Ding, ``Rate region of {ISAC} for
  pinching-antenna systems,'' \emph{arXiv preprint arXiv:2505.10179}, 2025.

\bibitem{zhang2025integrated}
Z.~Zhang, Z.~Wang, X.~Mu, B.~He, J.~Chen, and Y.~Liu, ``Integrated sensing and
  communications for pinching-antenna systems ({PASS}),'' \emph{arXiv preprint
  arXiv:2504.07709}, 2025.

\bibitem{mao2025multi}
W.~Mao, Y.~Lu, Y.~Xu, B.~Ai, O.~A. Dobre, and D.~Niyato, ``Multi-waveguide
  pinching antennas for {ISAC},'' \emph{arXiv preprint arXiv:2505.24307}, 2025.

\bibitem{wei2023integrated}
Z.~Wei, H.~Qu, Y.~Wang, X.~Yuan, H.~Wu, Y.~Du, K.~Han, N.~Zhang, and Z.~Feng,
  ``Integrated sensing and communication signals toward 5{G}-a and 6{G}: A
  survey,'' \emph{IEEE Internet Things J.}, vol.~10, no.~13, pp.
  11\,068--11\,092, 2023.

\bibitem{9540344}
J.~A. Zhang, F.~Liu, C.~Masouros, R.~W. Heath, Z.~Feng, L.~Zheng, and
  A.~Petropulu, ``An overview of signal processing techniques for joint
  communication and radar sensing,'' \emph{{IEEE} J. Sel. Top. Signal
  Process.}, vol.~15, no.~6, pp. 1295--1315, 2021.

\bibitem{liu2025pinching1}
Y.~Liu, H.~Jiang, X.~Xu, Z.~Wang, J.~Guo, C.~Ouyang, X.~Mu, Z.~Ding,
  A.~Nallanathan, G.~K. Karagiannidis \emph{et~al.}, ``Pinching-antenna systems
  ({PASS}): A tutorial,'' \emph{arXiv preprint arXiv:2508.07572}, 2025.

\bibitem{zhou2025channel}
G.~Zhou, V.~K. Papanikolaou, Z.~Ding, and R.~Schober, ``Channel estimation for
  mmwave pinching-antenna systems,'' in \emph{2025 IEEE 26th International
  Workshop on Signal Processing and Artificial Intelligence for Wireless
  Communications (SPAWC)}.\hskip 1em plus 0.5em minus 0.4em\relax IEEE, 2025,
  pp. 1--5.

\bibitem{grant2014cvx}
M.~Grant and S.~Boyd, ``{CVX}: Matlab software for disciplined convex
  programming, version 2.1,'' [Online]. Available:\url{http://cvxr.com/cvx},
  2014.

\bibitem{5447068}
Z.-Q. Luo, W.-K. Ma, A.~M.-C. So, Y.~Ye, and S.~Zhang, ``Semidefinite
  relaxation of quadratic optimization problems,'' \emph{{IEEE} Signal Process.
  Mag.}, vol.~27, no.~3, pp. 20--34, 2010.

\bibitem{9124713}
X.~Liu, T.~Huang, N.~Shlezinger, Y.~Liu, J.~Zhou, and Y.~C. Eldar, ``Joint
  transmit beamforming for multiuser {MIMO} communications and {MIMO} radar,''
  \emph{{IEEE} Trans. Signal Process.}, vol.~68, pp. 3929--3944, 2020.

\bibitem{boyd2004convex}
S.~P. Boyd and L.~Vandenberghe, \emph{Convex optimization}.\hskip 1em plus
  0.5em minus 0.4em\relax Cambridge university press, 2004.

\bibitem{10579914}
H.~Li, Z.~Wang, X.~Mu, P.~Zhiwen, and Y.~Liu, ``Near-field integrated sensing,
  positioning, and communication: A downlink and uplink framework,''
  \emph{{IEEE} J. Sel. Areas Commun.}, vol.~42, no.~9, pp. 2196--2212, 2024.

\bibitem{kay1993fundamentals}
S.~M. Kay, \emph{Fundamentals of statistical signal processing: estimation
  theory}.\hskip 1em plus 0.5em minus 0.4em\relax Prentice-Hall, Inc., 1993.

\end{thebibliography}

\end{document}